\documentclass[usenatbib]{mn2e}
\usepackage{epsfig}
\usepackage{amssymb}
\usepackage{lscape,graphicx}
\usepackage{longtable}
\def\850um{$850\,\mathrm{\mu m}$}

\def\350um{$350\,\mathrm{\mu m}$}

\usepackage{graphicx}
\usepackage{epstopdf}
\def\lsim{\mathrel{\lower2.5pt\vbox{\lineskip=0pt\baselineskip=0pt
           \hbox{$<$}\hbox{$\sim$}}}}

\def\gsim{\mathrel{\lower2.5pt\vbox{\lineskip=0pt\baselineskip=0pt
           \hbox{$>$}\hbox{$\sim$}}}}

\begin{document}

\title[SHARC-II follow-up of SHADES sources] {The SCUBA HAlf Degree Extragalactic Survey (SHADES) -- VI. $350\,\mathrm{\mu m}$ mapping of submillimetre galaxies}

\author[Coppin et al.]{ 
\parbox[t]{\textwidth}{
Kristen Coppin$^{1,2}$, Mark Halpern$^{1}$, Douglas Scott$^{1}$, Colin Borys$^{3}$, James Dunlop$^{4}$, Loretta Dunne$^{5}$, Rob Ivison$^{4,6}$, Jeff Wagg$^{7,8}$, Itziar Aretxaga$^{8}$, Elia Battistelli$^{1}$, Andrew Benson$^{9}$, Andrew Blain$^{9}$, Scott Chapman$^{9}$, Dave Clements$^{10}$, Simon Dye$^{11}$, Duncan Farrah$^{12}$, David Hughes$^{8}$, Tim Jenness$^{13}$, Eelco van Kampen$^{14}$, Cedric Lacey$^{2}$, Angela Mortier$^{4}$, Alexandra Pope$^{1}$, Robert Priddey$^{15}$, Stephen Serjeant$^{16}$, Ian Smail$^{2}$, Jason Stevens$^{15}$, Mattia Vaccari$^{17}$}\\\\
$^{1}$ Department of Physics \& Astronomy, University of British Columbia, 6224 Agricultural Road, Vancouver, British Columbia, V6T 1Z1, Canada\\
$^{2}$ Institute for Computational Cosmology, University of Durham, South Rd, Durham, DH1 3LE, UK\\
$^{3}$ Department of Astronomy \& Astrophysics, University of Toronto, 50 St. George Street, Toronto, Ontario, M5S 3H4, Canada\\
$^{4}$ Institute for Astronomy, University of Edinburgh, Royal Observatory, Blackford Hill, Edinburgh, EH9 3HJ, UK\\
$^{5}$ The Centre for Astronomy \& Particle Theory, The School of Physics \& Astronomy, University of Nottingham, University Park, Nottingham, NG7 2RD, UK\\ 
$^{6}$ UK ATC, Royal Observatory, Blackford Hill, Edinburgh, EH9 3HJ, UK\\
$^{7}$ National Radio Astronomy Observatory, P.O. Box 0, Socorro, NM, 87801, USA\\
$^{8}$ Instituto Nacional de Astrof\'{\i}sica, \'{O}ptica y Electr\'{o}nica, Apartado Postal 51 y 216, 72000 Puebla, Pue., Mexico\\
$^{9}$ Caltech, 1200 E. California Blvd, Pasadena, CA 91125-0001, USA\\
$^{10}$ Astrophysics Group, Blackett Laboratory, Imperial College, Prince Consort Rd., London SW7 2BW, UK\\
$^{11}$ School of Physics and Astronomy, Cardiff University, 5, The Parade, Cardiff CF24 3YB, UK\\ 
$^{12}$ Department of Astronomy, Cornell University, Space Sciences Building, Ithaca, NY 14853, USA\\
$^{13}$ Joint Astronomy Centre, 660 N.\ A`oh\={o}k\={u} Place, University Park, Hilo, HI 96720, USA \\
$^{14}$ Institute for Astro- and Particle Physics, University of Innsbruck, Technikerstr. 25, A-6020 Innsbruck, Austria\\
$^{15}$  Centre for Astrophysics Research, Science and Technology Research Institute, University of Hertfordshire, College Lane, Hatfield, Hertfordshire AL10 9AB, UK\\
$^{16}$ Department of Physics, The Open University, Milton Keynes, MK7 6AA, UK\\
$^{17}$ Department of Astronomy, University of Padova, Vicolo dell'Osservatorio 2, 35122, Italy\\
}

\maketitle

\begin{abstract}
  
  A follow-up survey using the Submillimetre High-Angular Resolution 
Camera (SHARC-II) at $350\,\mathrm{\mu m}$
has been carried out to map the regions around
  several $850\,\mathrm{\mu m}$-selected sources from the
  Submillimetre HAlf Degree Extragalactic Survey (SHADES).
  These observations probe the infrared
  luminosities and hence star-formation rates in the largest existing, most
  robust sample of submillimetre galaxies (SMGs).  We measure
  $350\,\mathrm{\mu m}$ flux densities for 24 $850\,\mathrm{\mu m}$
  sources, seven of which are detected at $\geq2.5\,\sigma$ within a 
10\,arcsec search radius of the $850\,\mathrm{\mu m}$ positions.
When results from the literature are included the total number 
of $350\,\mathrm{\mu m}$ flux density constraints
  of SHADES SMGs is 31, with 15 detections. 
  We fit a modified blackbody to the far-infrared (FIR)
  photometry of each SMG, and confirm that typical SMGs are dust-rich
  ($M_\mathrm{dust}\simeq9\times10^{8}\,\mathrm{M}_{\odot}$), luminous
  ($L_\mathrm{FIR}\simeq2\times10^{12}\,\mathrm{L}_{\odot}$)
  star-forming galaxies with intrinsic dust temperatures of
  $\simeq35\,\mathrm{K}$ and star-formation rates of
  $\simeq400\,\mathrm{M}_{\odot}\,\mathrm{yr}^{-1}$.  
  We have measured the temperature distribution of 
SMGs and find that the underlying distribution is 
slightly broader than implied by the error bars, 
and that most SMGs are at 28\,K with a few hotter.  We also
  place new constraints on the $350\,\mathrm{\mu m}$ source counts,
   $N_\mathrm{350}(>25\,\mathrm{mJy})\sim200$--$500\,\mathrm{deg}^{-2}$.

\end{abstract}

\begin{keywords} 
submillimetre -- surveys -- cosmology: observations -- 
galaxies: high-redshift -- galaxies: formation -- galaxies: starburst 
\end{keywords}

\section{Introduction}\label{intro}

The SCUBA HAlf Degree Extragalactic Survey (SHADES; \citealt{Mortier};
\citealt{Coppin06}) mapped $\simeq0.25\,\mathrm{deg^{2}}$ of sky with
an RMS of 2\,mJy at $850\,\mathrm{\mu m}$ with the Submillimetre
Common-User Bolometer Array (SCUBA; \citealt{Holland}).  The area was
split approximately evenly between the Lockman Hole (LH) and the
Subaru-\textit{XMM} Deep Field (SXDF).  Using uniform selection
criteria, the survey uncovered 120 SMGs with a median deboosted flux
density of $\sim\!5\,\mathrm{mJy}$ \citep{Coppin06}.  The SHADES
programme was designed to study the nature and evolution of high
star-formation rate (SFR) submillimetre galaxies (SMGs) via 
a systematic study of a
well-characterised and statistically meaningful sample. The programme
includes an effort to identify members of the source list at other
wavelengths using deep follow-up data from the radio \citep{paper3} to
X-ray, in order to characterise the SHADES population and to probe the
variation in the star-formation and clustering with redshift. The
relatively precise positions available from the radio data greatly aid
in identifying secure counterparts at other wavelengths, which can
then be used to provide spectroscopic or photometric redshifts
\citep{paper4} and to categorise the sources. 

Even combined with knowledge of source redshift, SCUBA
$850\,\mathrm{\mu m}$ and millimetre wavelength fluxes do not
constrain the total dust mass of a galaxy because there is an
ambiguity between column density and source temperature.  The
submillimetre (submm) spectral energy distribution (SED) 
of a luminous dusty galaxy arises from the
re-emission at far-infrared (FIR) wavelengths of absorbed optical/UV radiation from
regions of intense star-formation (see e.g.~\citealt{SandMir} and
references therein).  Typically the dust temperature is within a factor
of 2 of $T_\mathrm{d}\sim40\,\mathrm{K}$ \citep{Blain}, so the
restframe SED peaks in the range 60--$120\,\mathrm{\mu m}$ and the
SED is almost a simple power law at the SCUBA wavelengths and
longer. 

At a redshift of $\left\langle z\right\rangle\,{\sim}\,2$
or 3, typical of SMGs \citep{Chapman2005}, 
the peak of
the SED is shifted to be near $350\,\mathrm{\mu m}$, and 
the Submillimetre High Angular Resolution Camera (SHARC-II;
\citealt{Dowell}) at the Caltech Submillimetre Observatory (CSO)
is very well situated to provide the photometry needed to constrain
the temperatures, and therefore the luminosities and 
masses, of the SHADES sources.

We have therefore mapped a subset of the SHADES catalogue 
with SHARC-II.  In this paper we constrain the FIR SEDs of our sample
by fitting modified blackbody curves to the FIR photometry of 31
SHADES SMGs including SHARC-II data at $350\,\mathrm{\mu m}$.
Sections~\ref{obs} and \ref{dr} describe the observations and data
reduction.  Section~\ref{results} presents the $350\,\mathrm{\mu m}$
flux densities of the SHADES galaxies and the FIR SEDs.  
Section~\ref{discussion} 
provides a discussion of the results.  
Conclusions and future prospects are discussed in
Section~\ref{fp}.  We adopt the Cosmological parameters from the
\textit{WMAP} fits in \citet{Spergel}: $\Omega_\Lambda=0.73$,
$\Omega_\mathrm{m}=0.27$, and $H_\mathrm{0}=71$\,km\,s$^{-1}$\,Mpc$^{-1}$.

\section{Survey Design and Observations}\label{obs}

\subsection{Instrument Description}

SHARC-II is a background-limited 350 and $450\,\mathrm{\mu m}$
common-user continuum camera with a
$3\,\mathrm{arcmin}\times1.5\,\mathrm{arcmin}$ field of view (FOV) at
the 10.4-metre CSO in Hawaii \citep{Dowell}.  The dish has a very low surface error
($10.4\,\mathrm{\mu m}$ RMS at $350\,\mathrm{\mu m}$) due to the active
Dish Surface Optimization System \citep{Leong} which
corrects the primary for surface imperfections and gravitational
deformations as a function of elevation angle during observations to
improve the telescope efficiency and pointing.  The result is that the
CSO is probably the best telescope in the world at shorter
submm wavelengths.  The resulting beam size (with good
focus and pointing) is 9\,arcsec FWHM at $350\,\mathrm{\mu m}$.

\subsection{Sample Selection}

120 submm sources have been identified in the SHADES fields
\citep{Coppin06}. Obtaining useful photometry for the full set would
require several hundred hours of excellent weather, so we have chosen
to observe a carefully chosen subset of the SHADES catalogue.  Many
sources lie close enough together that multiple targets can be
selected within a single SHARC-II FOV.  We chose eight of our 14
fields to contain a large fraction of the SHADES close angular pairs.
This is mostly just for efficiency, but there is also a small chance that the
observations would help measure clustering in the SHADES catalogue. We
chose a deliberate mixture of sources with compact, extended, or no
radio counterparts \citep{paper3}: 11 sources with one robust
compact radio ID; three sources with two robust compact radio
counterparts; three sources with one extended-looking radio counterpart; and
seven sources with no reliable radio ID. Choosing
targets with this mixture of radio characteristics could
help refine the SHADES redshift distribution and test if there is a
sub-population with a high-redshift tail (e.g.~\citealt{Dunlop01},
\citealt{Ivison07}).

Two of our fields were also observed by \citet{Kovacs} or
\citet{Laurent06} (sources LOCK850.3, LOCK850.1, and LOCK850.41), which
allows for cross-calibration and checks of systematics in 
observation and analysis.  One field (LOCK44/45) covered 
two $850\,\mathrm{\mu m}$ source candidates from a preliminary 
SHADES catalogue which were later eliminated from the
official SHADES catalogue  since they were
deemed likely to be spurious sources \citep{Coppin06}.
This field is not used in the ensuing analysis.

\begin{table*}
\begin{minipage}{1.\textwidth}
\scriptsize
\caption[]{Summary of the SHARC-II observations and map properties.  RA and Dec. (J2000) coordinates are given for the field pointing centre, with each field containing one or more SHADES sources.  The date of observations, weather conditions (mean atmospheric opacity at $350\,\mathrm{\mu m}$), and the `effective' integration times (see Appendix~A; note that this is not the same as the actual time spent on the sky) are listed.  Since a large fraction of each map is comprised of noisy edges, we quote a median value of a `trimmed' noise map containing only pixels with more than 60 per cent of the maximum depth coverage achieved by the central pixel in the map.  Total map areas are also given.}\label{tab:sharc}
\begin{tabular}{llrrlcccc}
\hline 
\multicolumn{1}{l}{Field} & \multicolumn{1}{c}{SHADES} & \multicolumn{1}{c}{RA} & \multicolumn{1}{c}{Dec.} & \multicolumn{1}{c}{UT Date} & \multicolumn{1}{c}{Mean} & \multicolumn{1}{c}{Eff. Int. Time} & \multicolumn{1}{c}{Median RMS} & \multicolumn{1}{c}{Map Area}\\ 
\multicolumn{1}{l}{Name} & \multicolumn{1}{c}{Sources} & \multicolumn{1}{c}{(J2000)} & \multicolumn{1}{c}{(J2000)} & \multicolumn{1}{c}{(yy-mm-dd)} & \multicolumn{1}{c}{ $\tau_\mathrm{350\,\mu m}$} & \multicolumn{1}{c}{(min.)} & \multicolumn{1}{c}{(mJy)} & \multicolumn{1}{c}{($\mathrm{arcmin}^2$)}\\ 
\hline
LOCK21/28 & 21, 28 & \(10^{\mathrm{h}}52^{\mathrm{m}}57\mathrm{\fs}3\) & \(57^{\circ}30'54{\farcs}0\) & 05-02-28/-03-08,09 & 1.30 & 8.5 & 13 & 3.1  \\
LOCK26/32 & 26 & \(10^{\mathrm{h}}52^{\mathrm{m}}39{\mathrm{\fs}}7\) & \(57^{\circ}23'16{\farcs}4\) & 05-02-28 & 0.99 & 9.3 & 9 & 3.1 \\
LOCK4/69 & 4 & \(10^{\mathrm{h}}52^{\mathrm{m}}05{\mathrm{\fs}}5\) & \(57^{\circ}27'06{\farcs}4\) & 05-02-28 & 0.88 & 9.9 & 10 & 3.1 \\
LOCK3/47 & 3, 47 & \(10^{\mathrm{h}}52^{\mathrm{m}}37{\mathrm{\fs}}1\) & \(57^{\circ}24'57{\farcs}5\) & 05-02-28 & 0.88 & 1.1 & 20 & 3.0 \\
LOCK33/42 & 33, 77 & \(10^{\mathrm{h}}51^{\mathrm{m}}56{\mathrm{\fs}}5\) & \(57^{\circ}23'05{\farcs}5\) & 05-03-01 & 0.90 & 8.9 & 9 & 3.1 \\
LOCK10/48/64 & 10, 48, 64 & \(10^{\mathrm{h}}52^{\mathrm{m}}52{\mathrm{\fs}}0\) & \(57^{\circ}32'51{\farcs}4\) & 05-03-01 & 0.89 & 9.8 & 13 & 3.7 \\
LOCK1/41/63 & 1, 6, 41, 63 & \(10^{\mathrm{h}}51^{\mathrm{m}}57{\mathrm{\fs}}9\) & \(57^{\circ}24'60{\farcs}6\) & 05-03-01 & 0.80 & 13.7 & 23 & 3.9 \\
LOCK16/50 & 16 & \(10^{\mathrm{h}}51^{\mathrm{m}}49{\mathrm{\fs}}3\) & \(57^{\circ}26'44{\farcs}5\) & 05-03-08 & 1.33 & 7.5 & 17 & 3.8 \\
LOCK22/25 & 22 & \(10^{\mathrm{h}}51^{\mathrm{m}}35{\mathrm{\fs}}2\) & \(57^{\circ}33'26{\farcs}9\) & 05-03-08,09 & 1.52 & 6.0 & 17 & 3.5\\
LOCK44/45 & none & \(10^{\mathrm{h}}51^{\mathrm{m}}56{\mathrm{\fs}}8\) & \(57^{\circ}28'37{\farcs}9\) & 05-03-09 & 1.33 & 4.7 & 22 & 3.8 \\
LOCK15 & 15 & \(10^{\mathrm{h}}53^{\mathrm{m}}19{\mathrm{\fs}}1\) & \(57^{\circ}21'10{\farcs}5\) & 06-02-24 & 1.52 & 1.1 & 55 & 2.8 \\
SXDF1/11 & 1, 11 &  \(02^{\mathrm{h}}17^{\mathrm{m}}27{\mathrm{\fs}}9\) & \(-04^{\circ}59'38{\farcs}0\) & 04-08-30,31/-09-01 & 1.83 & 6.4 & 14 & 3.8 \\
SXDF3/8 & 3, 8 & \(02^{\mathrm{h}}17^{\mathrm{m}}43{\mathrm{\fs}}2\) & \(-04^{\circ}56'12{\farcs}5\) & 04-08-30,31/-09-01 & 1.81 & 5.2 & 15 & 4.3 \\
SXDF17/26 & 17, 119 & $02^{\mathrm{h}}17^{\mathrm{m}}55{\mathrm{\fs}}8$ & $-04^{\circ}52'50{\farcs}4$ & 05-03-01,03,05,07,08,09 & 1.39 & 6.8 & 17 & 3.1 \\
\end{tabular}
\end{minipage}
\end{table*}

\subsection{Observations}

Ten SHADES Lockman Hole (LH) fields were mapped over the course of
four superb stable weather nights in March 2005 and February 2006
($0.035\!<\!\tau_{\mathrm{225\,GHz}}\!<\!0.06$).  Four additional
fields were observed in more marginal observing conditions.  For each
field data were collected in 10 minute scans using a non-connecting
Lissajous scan pattern with small amplitudes of typically 20\,arcsec
in both altitude and azimuth and a `period' of 15--20 seconds.
Integration times, typical weather conditions and resulting map depths
of each SHARC-II field are given in Table~\ref{tab:sharc}.

By design, our strategy was to reach a  map RMS of
$3\sigma\simeq 30\,\mathrm{mJy}$ at $350\,\mathrm{\mu m}$, sufficient to detect or
place useful limits on an $S_{850}\sim8\,\mathrm{mJy}$ SMG with
$T_\mathrm{d}\simeq40$\,K at $z\lesssim3$.  Good weather is scarce, and
it is important to be as efficient as possible by only integrating
down to the planned noise level. We tracked the effective integrating
time on each source by compensating each 10 minute file for zenith
angle and atmospheric opacity at $350\,\mathrm{\mu m}$, as inferred
from the $\tau_{\mathrm{350\,\mu m}}$ and $\tau_{\mathrm{225\,GHz}}$ dippers. 
See Appendix~\ref{effective} for details.
Typically, 3--4
hours of observations were required for a map
$3\,\mathrm{arcmin}\times1.5\,\mathrm{arcmin}$ in size.

Pointing, focus checks, and calibration were performed hourly on
standard sources in close proximity to the science targets.  The same
scan pattern was used as for the science targets, but with integration
times of only 120--160 seconds (as typical flux densities are
$\gtrsim2\,\mathrm{Jy}$).

The typical pointing RMS of 2--3\,arcsec at the CSO is a
non-negligible fraction of the 9\,arcsec beamsize, and much care is
taken to minimise pointing errors.  Observations of point-like
galaxies, quasars, protostellar sources, HII regions, and evolved
stars are used by the CSO staff for constructing a pointing model.
The model predictions for the calibrators observed before and after
science observations are compared to the actual pointing
measurements; offsets are calculated and applied to the model pointing
predictions for the science observations during the map coaddition
stage of the reduction.  This procedure yields a pointing accuracy RMS
of $\sim2\,\mathrm{arcsec}$ (D. Dowell, private communication).

Flux calibration is performed by comparing the known and measured
flux densities and beamsizes obtained for standard calibration
sources.  For sources in the SXDF, oCeti (a Mira variable) and
occasionally OH231 (a proto-planetary nebula) were used as pointing
and calibration sources.  For sources in the LH, CIT6 (an evolved
star)  was used, and the nearby asteroids Pallas and Egeria
when CIT6 was unavailable.  The standards have well-tabulated
$350\,\mathrm{\mu m}$ flux densities, which are available from the
CSO/SHARC-II calibration web page\footnote{{\tt
    http://www.submm.caltech.edu/$\sim$sharc}}.  The final calibration
is expected to be better than 15 per cent, with systematic effects
being negligible at that level (CSO staff, private communication).

\section{Data Reduction}\label{dr}

\subsection{Map-Making and Source Extraction}

\setcounter{figure}{0}

We make maps from the raw data and then extract $350\,\mathrm{\mu m}$ fluxes
from the maps. 

\begin{figure*}
\epsfig{file=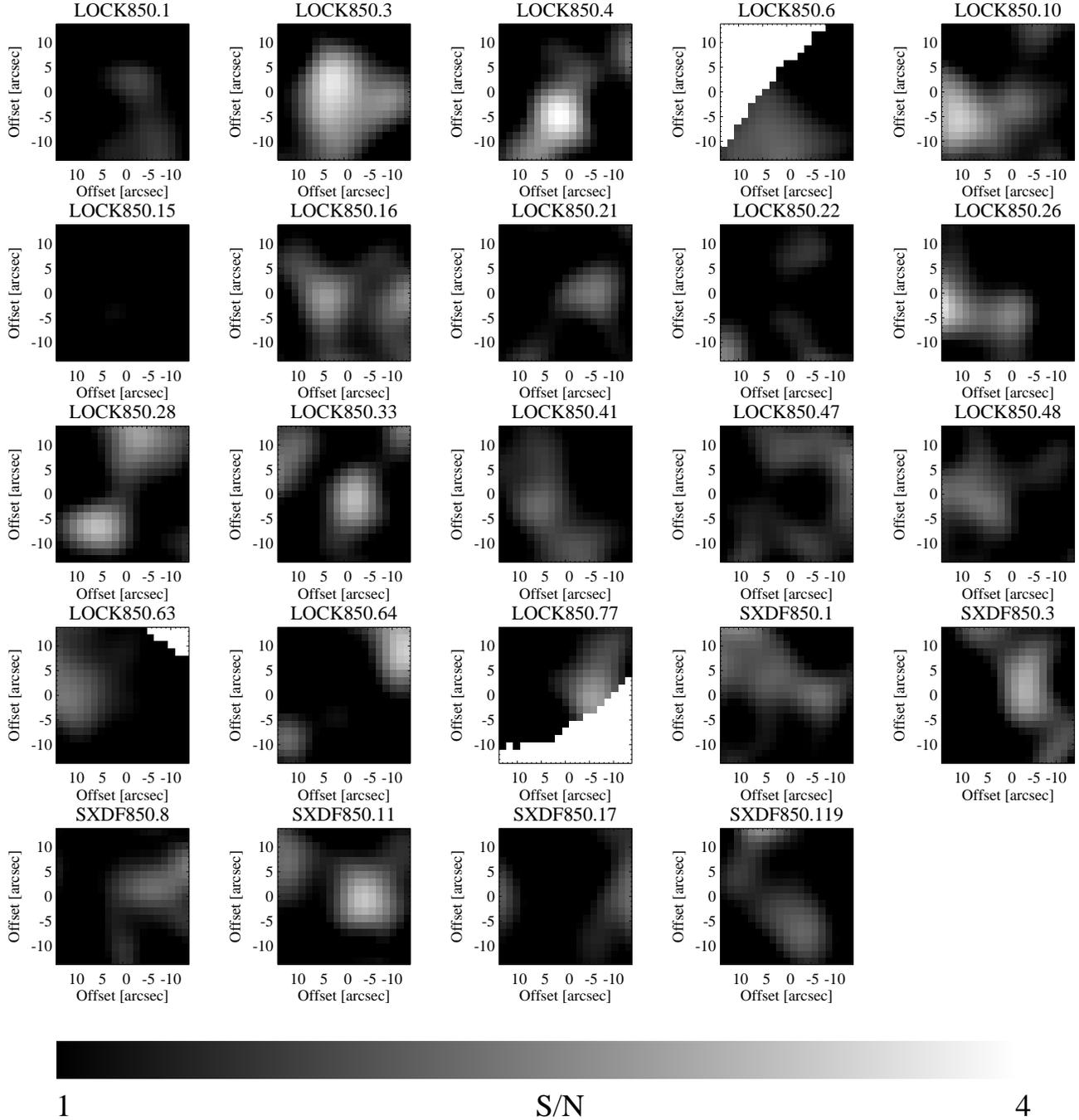,width=1.0\textwidth}
\caption[]{$30\times30$ $\mathrm{arcsec}^{2}$ SHARC-II
  $350\,\mathrm{\mu m}$ thumbnail images of each SHADES source,
  centred on the $850\,\mathrm{\mu m}$ positions and using
  1.62\,arcsec pixels.  Each thumbnail is extracted from a map that
  has been convolved with the FWHM=9\,arcsec 
SHARC-II point spread function.  The images are all displayed on
  the same flux density scale for comparison.  LOCK850.6, 63 and 77
  lie on the map edges.  See Table~\ref{tab:sharc_fluxes} for the
  corresponding flux densities, noise estimates, S/N values, and
  counterpart associations.}
\label{fig:thumb}
\end{figure*}

The map-making data reduction package is \small{\tt CRUSH} \normalsize
(Comprehensive Reduction Utility for SHARC-II), a Java-based tool
developed by \citet{Kovacsthesis}.  
The software iterates a least-squares algorithm
to solve for celestial emission along with instrumental
and atmospheric contributions to the total power signal.  \small{\tt CRUSH}
\normalsize  accesses the $\tau_{\mathrm{350\,\mu m}}$
polynomial fits to obtain a low noise  $350\,\mathrm{\mu m}$ sky
opacity-based estimate of  the atmospheric signal. Using
the `deep' utility, as is recommended for sources fainter than 
$100\,\mathrm{mJy}$, the
maps of each field are coadded on a grid of
$1.62\,\mathrm{arcsec}$ square pixels. The outer four
rows of pixels  are removed automatically by \small{\tt CRUSH}
\normalsize since
they are not sampled sufficiently well to converge to  
useful measurements.  The data are fitted with a single Gaussian
beam profile with a FWHM of 9\,arcsec.  Thumbnails of the reduced maps centred on the
SHADES $850\,\mathrm{\mu m}$ positions are shown in Fig.~\ref{fig:thumb}.
The bulk of the structure in these images is detector noise.

\subsection{Tests of the Data}\label{sharc_tests}
  
 Peaks are identified in maps of the signal-to-noise ratio (S/N) using an
algorithm which only accepts sources separated by at least
$3\,\sigma_{\mathrm{beam}}\!=\!11.5\,\mathrm{arcsec}$.  The maps are 
also inverted and negative sources are extracted in the same way as for the
positive map. The numbers of peaks found above a given threshold are
listed in Table \ref{tab:peaks}, as are the numbers of peaks expected
in a Gaussian field.   The total area of the maps searched for
peaks is $44.5\,\mathrm{arcmin}^2$.

\begin{table}
\begin{minipage}{1\textwidth}
\scriptsize
\caption{Numbers of positive and negative peaks in S/N maps.}
\label{tab:peaks}
\begin{tabular}{|c|r r r r r |}
\hline
\multicolumn{1}{c}{} & \multicolumn{1}{r}{$2\,\sigma$} & \multicolumn{1}{r}{$2.5\,\sigma$} & \multicolumn{1}{r}{$3\,\sigma$} & \multicolumn{1}{r}{$4\,\sigma$}\\\hline

Positive Peaks & 72 & 32 & 9~~ & 1 \\
 & & & & \\
\hline
Negative Peaks & 66 & 28 & 7~~ & 0 \\
$P_\mathrm{N}(10'')$  &12.9{\%} & 5.4\% & 1.4\% &--\\
\hline
Gaussian Model & 51 & 16 & 3.4 & 0.6 \\
$P_\mathrm{N}(10'')$    & 10.0\% & 3.1\% & 0.7\% &--\\
%
%
\hline
\end{tabular}
\end{minipage}
\end{table}

Notice that there are more negative peaks detected at any given noise
level than are expected under the assumption that the variance is what
one calculates from the time stream.  However, if the maps have
Gaussian distributed noise, but with a variance 7 per cent larger than we
infer from the time series the {\it Negative} and {\it
  Gaussian} columns in Table~\ref{tab:peaks} would match; the
number of Gaussian peaks expected at $S/N= 0.93\times( 2, 2.5, 3) $
is (67.0, 25.0, 6.3).  We find this level of agreement between variance
in the time series and variance in the maps encouraging, and use the
number of negative peaks to set the statistical significance of any
detections of positive flux in the maps.

\subsection{$350\,\mathrm{\mu m}$ flux densities of SHARC-II-detected SMGs}\label{sharc_flux}

If a statistically significant peak is found in the SHARC-II map within
a small search radius (discussed below) of a SHADES position, we 
infer that this is the
counterpart source and can use its flux, after deboosting as per
\citet{Coppin}, to constrain the source SED.
  
There are several different approaches to estimating the \350um\ flux
density associated with SHADES sources if a SHARC-II source is \textit{not}
detected, and they are biased in different ways.  
(1) Measuring the
flux density of each object at the \850um\ position will be biased low
on average because of the uncertainty in the true source location due
to the large beam sizes and modest S/N in the SHADES maps.  
(2)
Measuring the flux density of the brightest pixel within given radius
of the \850um\ position will be biased sightly high on
average\footnote{\citep{Kovacs} and \citealt{Laurent06} adopt this
  approach for measuring fluxes for detections and non-detections
  alike.} (e.g.~\citealt{Coppin}; see discussion below). 
(3) Measuring the $350\,\mathrm{\mu m}$
flux density of each object at the precise radio position, if one
exists, is less biased on average than methods 1 and 2 because of the
small positional uncertainty.

Not all of the
$850\,\mathrm{\mu m}$ sources mapped at $350\,\mathrm{\mu m}$ have
radio counterparts.  We therefore choose to measure the flux densities
using method 2 and to correct for flux boosting
effects, since this measurement can be performed for all of our target
SMGs in a uniform way.  Detections, and measurements from methods 1, 2 and 3
are given in Table~\ref{tab:sharc_fluxes}, without deboosting, for
completeness and inter-comparison. The deboosted fluxes are listed in
Table~\ref{tab:sharc_photom}. In fact we deboost the detected
sources and the other sources in the same way.

The first step in either source detection or method 2 is to choose a
search radius for companion sources. Too large a radius increases the
false detection rate and the flux boost factor.  Too small a radius causes
sources to be missed.  The rows labeled $P_\mathrm{N}(10'')$ in
Table~\ref{tab:peaks} show the percentage chance that an arbitrary
10\,arcsec circle contains a peak.  Using the negative peaks
as a reliable measurement of the noise level in the maps, we infer
that a $2.5\,\sigma$ peak found within 10\,arcsec of a SHADES
location is 5 per cent likely to be a false positive association, 
i.e.~a $2.5\,\sigma$ peak found within a 10\,arcsec radius is a 
95 per cent confidence detection.

We use Monte Carlo techniques on the actual data to test the number
of false identifications made. An area with a 10\,arcsec radius around
each $850\,\mathrm{\mu m}$ source is masked out so that any real
counterpart will not contaminate the test.  We select a random
position on each map and search for a peak in the map within the given
search radius above the designated S/N threshold.  This is repeated
10,000 times over the SHARC-II maps, and we find that using a search
radius of $\simeq\!10\,\mathrm{arcsec}$ and S/N\,$\gtrsim 2.5$ finds a
SHARC-II peak \textit{at random} in 5 per cent of the trials,
confirming the conclusion above from Table~\ref{tab:peaks}.  We
therefore adopt counterpart search criteria of $2.5\,\sigma$ and
10\,arcsec.

We can estimate the fraction of sources for which a $10\,\mathrm{arcsec}$
search misses the true companion
due to positional uncertainty in both the SHADES and SHARC-II data.
 \citet{paper3} find a one-dimensional
positional uncertainty of 3.2\,arcsec for SHADES SMGs by comparing the
$850\,\mathrm{\mu m}$ determined positions (beam
FWHM=$14.8\,\mathrm{arcsec}$) with more precise radio positions
(synthesised beam FWHM=$1.3\,\mathrm{arcsec}$).  This is consistent
with the theoretical expectation (see equation 2 of Appendix
B in \citealt{paper3}) of $\sigma\simeq
0.6\,\mathrm{(S/N)^{-1}\,FWHM}\simeq\!3$ for
FWHM=$14.8\,\mathrm{arcsec}$ and $\mathrm{S/N}\simeq3$. 
The positional uncertainty of SHARC-II observations stems from the
telescope pointing uncertainty, typically 2--3\,arcsec. Adding these
uncertainties in quadrature and converting one-dimensional
uncertainties to two dimensions and then a radial distance, we expect the
true \350um\ counterpart to lie within a 10\,arcsec circle 93 per cent of the
time.  Using a larger search radius would increase the level of flux
boosting without including more sources.  Peaks in the SHARC-II map will
be displaced from the true counterpart location by a distance that
scales with the SHARC-II beam size (9\,arcsec) and the S/N 
($\geq2.5\,\sigma$), as for SHADES. Adding this term, we find that
90\% of apparent peaks associated with a counterpart will lie within
our chosen 10\,arcsec circle.

Proper correction for the flux boosting that results from picking off
peaks in low S/N data requires knowledge of the source count
distribution (e.g.~\citealt{Coppin}).  However, prior information
about the \350um\ source counts is not yet sufficiently
well-constrained.  We have estimated the flux boosting effect on this
sample of low S/N $350\,\mathrm{\mu m}$-detected and non-detected SMGs
($\lesssim4\,\sigma$) by calculating the error-weighted mean excess
flux of method 2 compared to method 3 to be a factor of $\simeq1.5$.
We therefore divide each $350\,\mathrm{\mu m}$ flux measurement, from
method 2 (or from modest S/N detections), by 1.5 and present the
deboosted fluxes in Table~\ref{tab:sharc_photom} for use in
 fitting SEDs.  For the highest S/N $350\,\mathrm{\mu m}$-detected
SMGs ($\gtrsim4.5\,\sigma$; \citealt{Kovacs}), flux boosting effects
are negligible and so we do not correct these fluxes.

\subsection{Discussion of other $350\,\mathrm{\mu m}$ photometry of $850\,\mathrm{\mu m}$ SHADES sources}

The $350\,\mathrm{\mu m}$ flux densities have been constrained for 25
per cent of the SHADES $850\,\mathrm{\mu m}$ sources, with the current
work providing 21/31 of the SHADES $350\,\mathrm{\mu m}$ flux density
constraints.  Complementary SHARC-II observations of SMGs including
several SHADES sources have been performed by \citet{Laurent06} and
\citet{Kovacs}.  Since we include their photometry in the SED fits of
seven SHADES sources, it is worth describing how fluxes are measured
by  these groups.

\citet{Kovacs} conducted follow-up observations of bright
(${>}\,5\,\mathrm{mJy}$) radio-identified SCUBA sources with optical
spectroscopic redshifts, including LOCK850.3, LOCK850.14, LOCK850.18,
and LOCK850.30.  They detected 12/15 of their sample with a mean
S/N$\,{\simeq}\,4.5$ for the detections.

\citet{Laurent06} performed follow-up observations of 17 Bolocam
1.1\,mm-selected source candidates in the LH with SHARC-II.  Of the
17, ten are detected, including LOCK850.1, LOCK850.2, LOCK850.3,
LOCK850.12, LOCK850.14, LOCK850.41, LOCK850.27, and
LOCK850.76\footnote{LOCK850.3 and LOCK850.14 are detected by
  \citet{Kovacs} but also discussed by \citet{Laurent06}.} (see
\citealt{Laurent05} for counterpart
identifications\footnote{LOCK850.27 and LOCK850.76 were not previously
  associated with Bolocam sources, since they are new SHADES 
detections.  LOCK850.27 and LOCK850.76 lie 4.8 and 10.6\,arcsec away
  from two Bolocam sources, and so two new SCUBA/Bolocam associations
  are claimed here.}).

\citet{Kovacs} and \citet{Laurent06} quote as a flux the peak flux
density in an specified search radius around their target source
position, which is taken from either radio or mm data.  They do not
deboost these fluxes, but their observations are at lower noise
levels so this step is less important.

Three SHADES sources that we observe in this programme,
LOCK850.1, LOCK850.3 and LOCK850.41 have also been been observed by
\citet{Kovacs} and \citet{Laurent06}.  Because their observations are
at lower noise we use their fluxes in Table~\ref{tab:sharc_photom} and in
subsequent analysis.

\begin{table*}
\begin{minipage}{1.0\textwidth}
\scriptsize
\caption{
  SHARC-II $350\,\mathrm{\mu m}$ flux densities of $850\,\mathrm{\mu
    m}$ sources from SHADES.  Coordinates are only given for
  $\geq2.5\,\sigma$ SHARC-II detections $\leq10\,\mathrm{arcsec}$ away
  from the SHADES $850\,\mathrm{\mu m}$ position.  Each offset has an
  uncertainty of about $\pm2$\,arcsec, reflecting the uncertainty in
  the SHARC-II map astrometry due to the RMS pointing uncertainty.
  The flux densities in the `brightest pixel' column represent 
  the highest S/N pixel in
  the beam-convolved image within 10\,arcsec of the SHADES catalogue
  position - these fluxes are used in the SED fitting once 
  a correction has been made for flux boosting 
  (see Table~\ref{tab:sharc_photom} for the corrected fluxes). 
  Non-detections are given in parentheses; measurements
  were also made on the SHARC-II maps at the SHADES $850\,\mathrm{\mu
    m}$ and radio positions for comparison.  See the text for a
  discussion of the biases of each of these measurements.  
  The observations for LOCK850.15 are insufficiently deep (incomplete)
  and its map is thus left out of the general analysis (see text).}
\label{tab:sharc_fluxes}
\begin{tabular}{lcccllll}
\hline 
\multicolumn{1}{c}{SHADES ID} & \multicolumn{2}{c}{$350\,\mathrm{\mu m}$ position} & \multicolumn{1}{c}{Offset} & \multicolumn{3}{c}{$S_{350}$} & \multicolumn{1}{l}{Comments} \\
\hline
\multicolumn{1}{c}{} & \multicolumn{1}{c}{RA} & \multicolumn{1}{c}{Dec.} & \multicolumn{1}{c}{} & \multicolumn{1}{c}{\textbf{brightest pixel ($\leq10''$)}} \vline & \multicolumn{1}{c}{at $850\,\mathrm{\mu m}$ posn.} \vline & \multicolumn{1}{c}{at radio posn.} & \multicolumn{1}{c}{} \\
\hline
\multicolumn{1}{c}{} & \multicolumn{1}{c}{(J2000)} & \multicolumn{1}{c}{(J2000)} & \multicolumn{1}{c}{(arcsec)} & \multicolumn{1}{c}{\textbf{(mJy)}} & \multicolumn{1}{c}{(mJy)} & \multicolumn{1}{c}{(mJy)} & \multicolumn{1}{c}{} \\
\hline
LOCK850.01 & -- & -- & ($\simeq4.3$) & ($23.0\,\pm{21.2}$) & $12.7\,\pm{22.2}$ & $22.6\,\pm{21.4}$ & see \citet{Laurent06} \\
LOCK850.03 & \(10^{\mathrm{h}}52^{\mathrm{m}}38{\mathrm{\fs}}70\) & \(57^{\circ}24'37{\farcs}4\) & 3.7 & $67.1\,\pm{18.5}$ ($3.6\,\sigma$) & $59.7\,\pm{18.0}$ & $65.0\,\pm{18.2}$ & see \citet{Kovacs} \\
 & & & & & & $58.2\,\pm{18.0}$ & \\
LOCK850.04 & \(10^{\mathrm{h}}52^{\mathrm{m}}04{\mathrm{\fs}}42\) & \(57^{\circ}26'54{\farcs}3\) & 5.0 & $37.3\,\pm{9.1}$ ($4.1\,\sigma$) & $21.0\,\pm{9.2}$ & $18.9\,\pm{9.3}$ & \\
 & & & & & & $33.2\,\pm{9.1}$ & \\
LOCK850.06 & -- & -- & ($\simeq10$) & ($57.0\,\pm{37.6}$) & $1.8\,\pm{43.1}$ & $23.6\,\pm{39.1}$ & edge of SHARC-II map \\
LOCK850.10 & -- & -- & ($\simeq10$) & ($36.2\,\pm{12.0}$) & $14.9\,\pm{11.9}$ & $19.4\,\pm{12.2}$ & \\
LOCK850.15 & -- & -- & ($\simeq4.2$) & ($9.8\,\pm{55.7}$) & $-32.7\,\pm{53.7}$ & $-8.3\,\pm{52.5}$ & incomplete observations \\
 & & & & & & $6.9\,\pm{54.9}$ & \\
LOCK850.16 & -- & -- & ($\simeq4.9$) & ($38.6\,\pm{15.8}$) & $26.6\,\pm{15.5}$ & $33.3\,\pm{15.6}$ & \\
LOCK850.21 & -- & -- & ($\simeq6.5$) & ($25.1\,\pm{13.2}$) & $14.3\,\pm{12.9}$ & -- & \\
LOCK850.22 & -- & -- & ($\simeq10$) & ($13.1\,\pm{15.6}$) & $-1.3\,\pm{15.6}$ & -- & \\
LOCK850.26 & -- & -- & ($\simeq4.5$) & ($18.3\,\pm{8.8}$) & $6.9\,\pm{8.5}$ & $13.4\,\pm{8.6}$ & \\
LOCK850.28 & \(10^{\mathrm{h}}52^{\mathrm{m}}57{\mathrm{\fs}}86\) & \(57^{\circ}30'59{\farcs}7\) & 10.0 & $34.9\,\pm{11.7}$ ($3.0\,\sigma$) & $8.3\,\pm{11.8}$ & ($29.6\,\pm{11.6}$) & \\
LOCK850.33 & \(10^{\mathrm{h}}51^{\mathrm{m}}55{\mathrm{\fs}}82\) & \(57^{\circ}23'11{\farcs}3\) & 1.4 & $24.7\,\pm{8.4}$ ($2.9\,\sigma$) & $21.5\,\pm{8.4}$ & $20.4\,\pm{8.4}$ & \\
LOCK850.41 & -- & -- & ($\simeq7.5$) & ($42.5\,\pm{25.4}$) & $15.9\,\pm{24.4}$ & $12.4\,\pm{24.2}$ & see \citet{Laurent06} \\
 & & & & & & $34.8\,\pm{24.8}$ & \\
LOCK850.47 & -- & -- & ($\simeq9.4$) & ($24.4\,\pm{20.9}$) & $-8.3\,\pm{21.2}$ & -- & \\
LOCK850.48 & -- & -- & ($\simeq6.5$) & ($24.3\,\pm{13.5}$) & $9.9\,\pm{12.3}$ & -- & \\
LOCK850.63 & -- & -- & ($\simeq10$) & ($52.3\,\pm{29.1}$) & $9.9\,\pm{31.4}$ & $15.6\,\pm{29.1}$ & \\
LOCK850.64 & -- & -- & ($\simeq10$) & ($17.4\,\pm{12.4}$) & $-5.1\,\pm{12.8}$ & -- & \\
LOCK850.77 & \(10^{\mathrm{h}}51^{\mathrm{m}}56{\mathrm{\fs}}22\) & \(57^{\circ}22'09{\farcs}8\) & 6.3 & $62.1\,\pm{24.6}$ ($2.5\,\sigma$) & $9.9\,\pm{18.7}$ & $-0.2\,\pm{17.9}$ & edge of SHARC-II map \\
SXDF850.1 & -- & -- & ($\simeq6.2$) & ($24.5\,\pm{13.9}$) & $15.9\,\pm{14.5}$ & $15.9\,\pm{14.6}$ & \\
SXDF850.3 & \(02^{\mathrm{h}}17^{\mathrm{m}}41{\mathrm{\fs}}95\) & \(-04^{\circ}56'26{\farcs}3\) & 3.4 & $39.3\,\pm{14.3}$ ($2.7\,\sigma$) & $23.1\,\pm{14.1}$ & $23.1\,\pm{14.1}$ & \\
SXDF850.8 & -- & -- & ($\simeq6.3$) & ($24.5\,\pm{14.1}$) & $10.5\,\pm{14.1}$ & $18.1\,\pm{14.5}$ & \\
SXDF850.11 & \(02^{\mathrm{h}}17^{\mathrm{m}}24{\mathrm{\fs}}81\) & \(-04^{\circ}59'37{\farcs}2\) & 4.6 & $46.6\,\pm{15.0}$ ($3.1\,\sigma$) & $26.6\,\pm{14.2}$ & $22.2\,\pm{14.1}$ & \\
SXDF850.17 & -- & -- & ($\simeq10$) & ($18.6\,\pm{21.0}$) & $-13.0\,\pm{17.3}$ & -- & \\
SXDF850.119 & -- & -- &($\simeq7.6$) & ($23.8\,\pm{14.9}$) & $13.0\,\pm{15.3}$ & $0.8\,\pm{15.4}$ & \\
\hline
\end{tabular}
\end{minipage}
\end{table*}
\normalsize

\subsection{Formally detected \350um\ counterparts}

In the analysis section of this paper we use our best estimates of
the $350\,\mathrm{\mu m}$ flux of each source as photometric constraints.
However, for comparison to blank field surveys it is useful to
consider how many of the sources are seen at a high enough
S/N that they would be counted as detections.

We have identified seven $350\,\mathrm{\mu m}$ counterparts of
$850\,\mathrm{\mu m}$ sources, as presented in
Table~\ref{tab:sharc_fluxes}.  The positional offsets of the
detections from the SHADES positions are consistent with the
positional distribution discussed above; 5/7 (71 per cent) of the
detections lie within 5\,arcsec.  No trend is seen of decreasing
positional offset with increasing S/N ratio of the seven associations,
but this is not surprising given the low number statistics and the
small dynamic range in S/N.

It is striking that the number of {\it detected} counterparts to SHADES 
sources exactly matches the excess of positive peaks over noise
(or negative peaks) in the full mapped area, 
as listed in Table~\ref{tab:peaks}.

As we did with the number of peaks, we can use the corresponding
composite negative catalogue to test if real associations are being
found or just noisy peaks in the $350\,\mathrm{\mu m}$ maps.
Counterparts to $850\,\mathrm{\mu m}$ sources are searched for in an
inverted map in the same way, using a S/N threshold of 2.5 and search
radius of 10\,arcsec and no `negative' counterparts are identified.
This result reassures us that the 350/$850\,\mathrm{\mu m}$
associations are likely to be real.

\subsection{$350\,\mathrm{\mu m}$ flux density constraints of SMGs not 
individually detected by SHARC-II and flux boosting}\label{deboosting}

Any of the following mechanisms might cause an SMG not to be detected
at $350\,\mathrm{\mu m}$:
(1) the $850\,\mathrm{\mu m}$ source could be
intrinsically less luminous, colder and/or have a different SED shape
than the strongly detected sources;
(2) the SMG could lie at a
redshift in excess of about 3, this being more likely if there is also
no radio counterpart, since the existing radio data are sufficiently
deep to obtain a large fraction of counterparts for SMGs only for
$z\lesssim 3$ (see \citealt{Ivison05}), and above $z\sim 3$, the
$350\,\mathrm{\mu m}$ band is sampling the Wien side of the SED and
suffers from cosmological dimming without the benefit of the negative
K-correction, or (3) 
The SMG could be spurious, although the false positive rate in the 
SHADES catalogue is believed to be very low. 
The catalogue was carefully constructed
from a comparison of multiple data reductions in order to minimise
false detections (see \citealt{Coppin06}).

We rely on the photometric fluxes even for the 16 sources which are
not detected at high significance.  Therefore we have
performed a stacking analysis to determine if a significantly
positive $350\,\mathrm{\mu m}$ flux density is associated
with those sources.

$350\,\mathrm{\mu m}$ flux densities and errors
are measured on the SHARC-II maps at the radio position for the 11
of these 16 sources which have a radio counterpart and we find the
mean flux is $16.7 \pm 4.8$\,mJy.  This is our least biased flux
estimator, and the agreement with our listed deboosted fluxes is very
encouraging.  The weighted mean deboosted flux for all 16 of these low
significance sources is $16.6 \pm 3.8$\,mJy.  

\begin{table*}
\begin{minipage}{1.0\textwidth}
\scriptsize
\begin{tabular}{lccccccccc}
\hline 
\multicolumn{1}{c}{SHADES ID} & \multicolumn{1}{c}{$S_{350}$} & \multicolumn{1}{c}{$S_{850}$} & \multicolumn{1}{c}{$S_{450}$} & \multicolumn{1}{c}{$S_{1.4\,\mathrm{GHz}}$} & \multicolumn{1}{c}{$S_{24}$} & \multicolumn{1}{c}{$S_{1.1\,\mathrm{mm}}$} & \multicolumn{1}{c}{$S_{1.2\,\mathrm{mm}}$} & \multicolumn{1}{c}{phot $z$} & \multicolumn{1}{c}{spec $z$} \\
\multicolumn{1}{c}{} & \multicolumn{1}{c}{(mJy)} & \multicolumn{1}{c}{(mJy)} & \multicolumn{1}{c}{($\mu \mathrm{Jy}$)} & \multicolumn{1}{c}{($\mu \mathrm{Jy}$)} & \multicolumn{1}{c}{(mJy)} & \multicolumn{1}{c}{(mJy)} & \multicolumn{1}{c}{} & \multicolumn{1}{c}{} \\
\hline
LOCK850.04 & $24.9\,\pm{9.1}$ & 10.65 ($\pm^{1.7}_{1.8}$) & $<134$ & $32.0\,\pm{5.1}$ & $261\,\pm{73}$ & -- & $3.7\,\pm{0.4}$ & 1.6 ($\pm^{0.3}_{0.1}$) & $(0.526\,\mathrm{or}\,1.482)^{2}$ \\
 & &  & & $73.0\,\pm{5.0}$ & $179\,\pm{68}$ & & & & \\
LOCK850.06 & $38.0\,\pm{37.6}$ & 6.85 ($\pm^{1.3}_{1.3}$) & $<77$ & $15.0\,\pm{4.8}$ (tent.) & $75.1\,\pm{12.7}$ & -- & -- & 3.6 ($\pm^{1.0}_{0.1}$) & -- \\
LOCK850.10 & $24.1\,\pm{12.0}$ & 9.15 ($\pm^{2.7}_{2.9}$) & $<365$ & $25.5\,\pm{6.3}$ & -- & -- & -- & 3.1 ($\pm^{0.9}_{0.3}$) & -- \\
LOCK850.15 & $6.5\,\pm{55.7}$ & 13.25 ($\pm^{4.3}_{5.0}$) & $<149$ & $43.9\,\pm{7.8}$ & $353\,\pm{20}$ & -- & $4.1\,\pm{0.7}$ & 2.4 ($\pm^{0.4}_{0.4}$) & -- \\
 & & & & $61.5\,\pm{7.6}$ & & & & & \\
LOCK850.16 & $25.7\,\pm{15.8}$ & 5.85 ($\pm^{1.8}_{1.9}$) & $<67$ & $106.0\,\pm{6.0}$ & $314\,\pm{24}$ & -- & $1.8\,\pm{0.5}$ & 1.9 ($\pm^{0.4}_{0.1}$) & $1.147^{2}$\\
LOCK850.21 & $16.7\,\pm{13.2}$ & 4.15 ($\pm^{2.0}_{2.5}$) & $<70$ & $5\,\sigma<30$ & $97.9\,\pm{14.1}$ & -- & $1.6\,\pm{0.4}$ & $\geq1.0$ & -- \\
LOCK850.22 & $8.7\,\pm{15.6}$ & 7.55 ($\pm^{3.2}_{4.2}$) & $<76$ & $5\,\sigma<30$ & $402\,\pm{21}$ & -- & -- & $\geq2.0$ & -- \\
LOCK850.26 & $12.2\,\pm{8.8}$ & 5.85 ($\pm^{2.4}_{2.9}$) & $<48$ & $31.4\,\pm{5.2}$ & $195\,\pm{16}$ & -- & -- & 3.6 ($\pm^{0.1}_{0.8}$) & -- \\
LOCK850.28 & $23.3\,\pm{11.7}$ & 6.45 ($\pm^{1.7}_{1.8}$) & $<56$ & -- & -- & -- & -- & $\geq2.0$ & -- \\
LOCK850.33 & $16.5\,\pm{8.4}$ & 3.85 ($\pm^{1.0}_{1.1}$) & $<49$ & $51.0\,\pm{4.3}$ & -- & -- & $2.8\,\pm{0.6}$ & 3.6 ($\pm^{0.7}_{0.9}$) & $2.664^{2},2.686^{1},3.699^{3}$ \\
LOCK850.47 & $16.3\,\pm{20.9}$ & 3.55 ($\pm^{1.7}_{2.1}$) & $<21$ & -- & -- & -- & -- & $\geq1.5$ & -- \\
LOCK850.48 & $16.2\,\pm{13.5}$ & 5.45 ($\pm^{2.1}_{2.5}$) & $<79$ & -- & $203\,\pm{17}$ & -- & $1.6\,\pm{0.4}$ & 2.4 ($\pm^{0.5}_{0.1}$) & -- \\
LOCK850.63 & $34.9\,\pm{29.1}$ & 3.65 ($\pm^{1.2}_{1.3}$) & $<50$ & $22.6\,\pm{4.8}$ & $236\,\pm{17}$ & -- & -- & 2.6 ($\pm^{0.4}_{0.4}$) & -- \\
LOCK850.64 & $11.6\,\pm{12.4}$ & 5.85 ($\pm^{2.5}_{3.2}$) & $<95$ & -- & -- & -- & $1.7\,\pm{0.4}$ & $\geq1.5$ & -- \\
LOCK850.77 & $41.4\,\pm{24.6}$ & 3.25 ($\pm^{1.2}_{1.3}$) & $<39$ & $15.5\,\pm{4.4}$ (tent.) & $51.7\,\pm{13.1}$ & -- & -- & 2.6 ($\pm^{0.8}_{0.1}$) & -- \\
SXDF850.1 & $16.3\,\pm{13.9}$ & 10.45 ($\pm^{1.5}_{1.4}$) & $<65$ & $54.3\,\pm{9.7}$ & -- & -- & -- & 2.6 ($\pm^{0.4}_{0.3}$)& -- \\
SXDF850.3 & $26.2\,\pm{14.3}$ & 8.75 ($\pm^{1.5}_{1.6}$) & $<81$ & $77.2\,\pm{9.3}$ & no ID & -- & -- & 2.1 ($\pm^{0.3}_{0.1}$) & -- \\
SXDF850.8 & $16.3\,\pm{14.1}$ & 6.05 ($\pm^{1.8}_{1.9}$) & $<45$ & $52.0\,\pm{9.5}$ & -- & -- & -- & 2.6 ($\pm^{1.3}_{0.1}$) & -- \\
SXDF850.11 & $31.1\,\pm{15.0}$ & 4.55 ($\pm^{1.9}_{2.2}$) & $<79$ & $56.8\,\pm{10.0}$ & $195\,\pm{47}$ & -- & -- & 2.4 ($\pm^{0.4}_{0.4}$) & -- \\
SXDF850.17 & $12.4\,\pm{21.0}$ & 7.65 ($\pm^{1.7}_{1.7}$) & $<71$ & -- & -- & -- & -- & $\geq2.0$ & -- \\
SXDF850.119 & $15.9\,\pm{14.9}$ & 4.55 ($\pm^{2.1}_{2.5}$) & $<70$ & $38.0\,\pm{9.7}$ (tent.) & $784\,\pm{47}$ & -- & -- & 2.1 ($\pm^{0.1}_{0.1}$) & -- \\
 & & & & & $275\,\pm{47}$ & & & & \\
\hline
LOCK850.01 & $24.1\,\pm{5.5}$ & 8.85 ($\pm^{1.0}_{1.0}$) & $<47$ & $78.9\,\pm{4.7}$ & $217\,\pm{16}$ & $4.4\,\pm{1.3}$ & $3.6\,\pm{0.5}$ & 2.4 ($\pm^{1.1}_{0.2}$) & $2.148^{1,2}$ \\
LOCK850.02 & $25.3\,\pm{10.3}$ & 13.45 ($\pm^{2.1}_{2.1}$) & $<123$ & $40.7\,\pm{5.6}$ & $545\,\pm{31}$ & $6.8\,\pm{1.4}$ & $5.7\,\pm{1.0}$ & 2.9 ($\pm^{0.3}_{0.1}$) & -- \\
 & & & & $52.4\,\pm{5.6}$ & & & & 2.9 ($\pm^{0.7}_{0.1}$) & \\
LOCK850.03 & $40.5\,\pm{6.5}$ & 10.95 ($\pm^{1.8}_{1.9}$) & $<34$ & $35.0\,\pm{5.2}$ & $183\,\pm{33}$ & $4.8\,\pm{1.3}$ & $4.6\,\pm{0.4}$ & 2.6 ($\pm^{0.3}_{0.1}$) & $(2.94\,\mathrm{or}\,3.036^{1})^{2}$\\
 & & & & $25.8\,\pm{4.9}$ & $175\,\pm{23}$ & & & & \\
LOCK850.12 & $29.3\,\pm{16.0}$ & 6.15 ($\pm^{1.7}_{1.7}$) & $<35$ & $44.3\,\pm{5.1}$ & $263\,\pm19$ & $4.1\,\pm{1.3}$ & $2.6\,\pm{0.4}$ & 2.6 ($\pm^{0.2}_{0.1}$) & $2.142^{1,2}$ \\
LOCK850.14 & $41.0\,\pm{6.8}$ & 7.25 ($\pm^{1.8}_{1.9}$) & $<96$ & -- & -- & $5.1\,\pm{1.3}$ & $3.4\,\pm{0.6}$ & 2.6 ($\pm^{0.8}_{0.1}$) & $2.611^{1,2,3}$ \\
LOCK850.18 & $7.5\,\pm{6.7}$ & 6.05 ($\pm^{1.9}_{2.1}$) & $<84$ & $29.4\,\pm{4.4}$ & -- & $5.1\,\pm{1.3}$ & $3.4\,\pm{0.6}$ & 3.1 ($\pm^{2.9}_{0.1}$) & $1.956^{1,2}$ \\
LOCK850.27 & $3.4\,\pm{5.1}$ & 5.05 ($\pm^{1.3}_{1.3}$) & $<32$ & -- & -- & $5.2\,\pm{1.4}$ & $3.2\,\pm{0.7}$ & 4.6 ($\pm^{1.4}_{0.4}$) & -- \\
LOCK850.30 & $38.0\,\pm{7.2}$ & 4.75 ($\pm^{1.5}_{1.6}$) & $<86$ & $245\,\pm{13}$ & $233\,\pm{19}$ & -- & ($0.4\,\pm{0.8}$) & 2.1 ($\pm^{0.1}_{0.4}$) & $2.689^{1,3}$ \\
LOCK850.41 & $10.3\,\pm{5.5}$ & 3.85 ($\pm^{0.9}_{1.0}$) & $<16$ & $43.6\,\pm{4.7}$ & $651\,\pm{46}$ & $4.0\,\pm{1.3}$ & $2.4\,\pm{0.5}$ & 3.4 ($\pm^{0.7}_{0.2}$) & $0.689^{1,2,4}$ \\
 & & & & $22.1\,\pm{4.8}$ & $475\,\pm{37}$ & & & & \\
LOCK850.76 & $4.4\,\pm{6.7}$ & 4.75 ($\pm^{2.5}_{3.1}$) & $<90$ & $48.0\,\pm{6.0}$ & $592\,\pm{26}$ & $4.4\,\pm{1.4}$ & -- & 4.6 ($\pm^{1.4}_{1.1}$) & -- \\
\hline
\end{tabular}
\caption{Multi-wavelength photometry of SHARC-II-observed SHADES sources.  The
  first set are SHADES sources that use $350\,\mathrm{\mu m}$ flux
  densities measured here, and the second set are $350\,\mathrm{\mu
    m}$ flux densities of SHADES sources obtained from the literature
  (\citealt{Kovacs} and \citealt{Laurent06}).  $350\,\mathrm{\mu m}$
  photometry has been corrected for flux boosting (see
  Section~\ref{deboosting}).  850 (deboosted) and $450\,\mathrm{\mu
    m}$ photometry are from \citet{Coppin06}, $1.4\,\mathrm{GHz}$ and
  $24\,\mathrm{\mu m}$ data are from \citet{paper3} (`tent.' refers to
  tentative identifications), 1.1\,mm and 1.2\,mm photometry are from
  \citet{Laurent05} and Dunlop (private communication; an improved
  reduction of the \citealt{Greve} MAMBO data), respectively.  In
  cases where two robust radio and/or $24\,\mathrm{\mu m}$ IDs exist for
  a single SHADES source, both are listed.  Photometric redshifts are
  from tables~3, 4 and 5 of \citet{paper4} and spectroscopic redshifts
  are from: 1.~\citet{Chapman2005}; 2.~\citet{Ivison05};
  3.~\citet{Chapman2003}; 4.~\citet{Chapman2002}.}
\label{tab:sharc_photom}
\end{minipage}
\end{table*}
\normalsize

\begin{figure}
\psfig{file=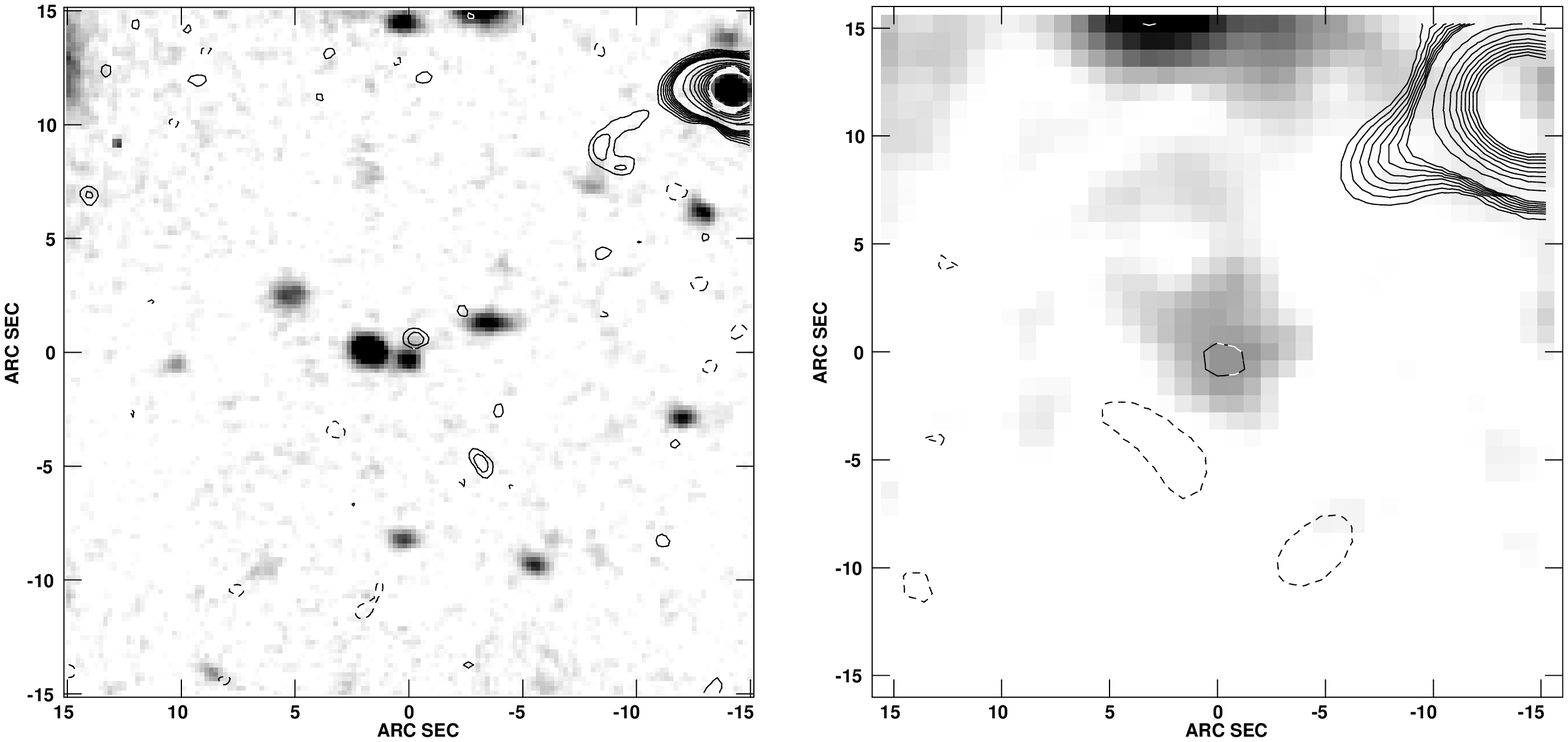,width=0.5\textwidth}
\caption[]
{A multi-wavelength view of LOCK350.1, a $3.9\,\sigma$
  $350\,\mathrm{\mu m}$ blank-field source detected at
  RA=\(10^{\mathrm{h}}52^{\mathrm{m}}43{\fs}2\),
  Dec.=\(57^{\circ}23'9''\) (J2000).  The
  $30\,\mathrm{arcsec}\times30\,\mathrm{arcsec}$ images are centred on
  the position of LOCK350.1.  \textit{Left:} High-resolution
  ($1.3\,\mathrm{arcsec}$) 1.4\,GHz contours ($-3$, 3, 4...10,
  20...$100\,\times\,\sigma$) from \citet{IvisonBiggs06} overlaid on
  greyscale archival $R$-band optical data.  \textit{Right:}
  Low-resolution ($4.2\,\mathrm{arcsec}$) 1.4\,GHz contours ($-3$, 3,
  4...10, 20...$100\,\times\,\sigma$) from \citet{IvisonBiggs06}
  overlaid on greyscale $24\,\mathrm{\mu m}$ data from \citet{paper3}.
}
\label{fig:lock350.1}
\end{figure}

\subsection{Serendipitous $350\,\mathrm{\mu m}$ blank-field SMGs}

We identify one serendipitous $3.9\,\sigma$ $350\,\mathrm{\mu m}$
source in our map of LOCK26/32: LOCK350.1.  LOCK350.1 is not
associated with any $850\,\mathrm{\mu m}$ sources, since it is
$\gtrsim15\,\mathrm{arcsec}$ away from any of the SHADES
$850\,\mathrm{\mu m}$ positions in the map.  Based on Gaussian noise
statistics and the number of independent beam sizes in the survey area
with noise less than 20\,mJy\footnote{A noise cut of 20\,mJy was
  chosen arbitrarily, since higher S/N detections in noisier regions
  are more likely to be spurious; e.g.~\citet{Coppin}.}, about 0.1
false positives are expected on average at a S/N of 3.9; LOCK350.1 is
therefore likely a real blank-field $350\,\mathrm{\mu m}$ source.  The
position and flux density of LOCK350.1 are
RA=\(10^{\mathrm{h}}52^{\mathrm{m}}43{\fs}2\),
Dec.=\(57^{\circ}23'09''\) (J2000) and $32.8\,\pm{8.3}\,\mathrm{mJy}$.  
This position corresponds to a region of positive flux
density in the $850\,\mathrm{\mu m}$ map and a $2.3\,\sigma$ peak is
located $\simeq4\,\mathrm{arcsec}$ from the position of LOCK350.1.
There also appears to be a $4\,\sigma$ 1.4\,GHz source and a faint
$24\,\mathrm{\mu m}$ counterpart at these coordinates
(\citealt{paper3}; \citealt{IvisonBiggs06}; see
Fig.~\ref{fig:lock350.1}).  This source is therefore probably the
third secure $350\,\mathrm{\mu m}$ blank-field detection (see
\citealt{Khan05} and \citealt{Khan} for the first two).

\section{Analysis and Results}\label{results}

\subsection{Constraints on the $350\,\mathrm{\mu m}$ source counts}

\begin{figure}
\epsfig{file=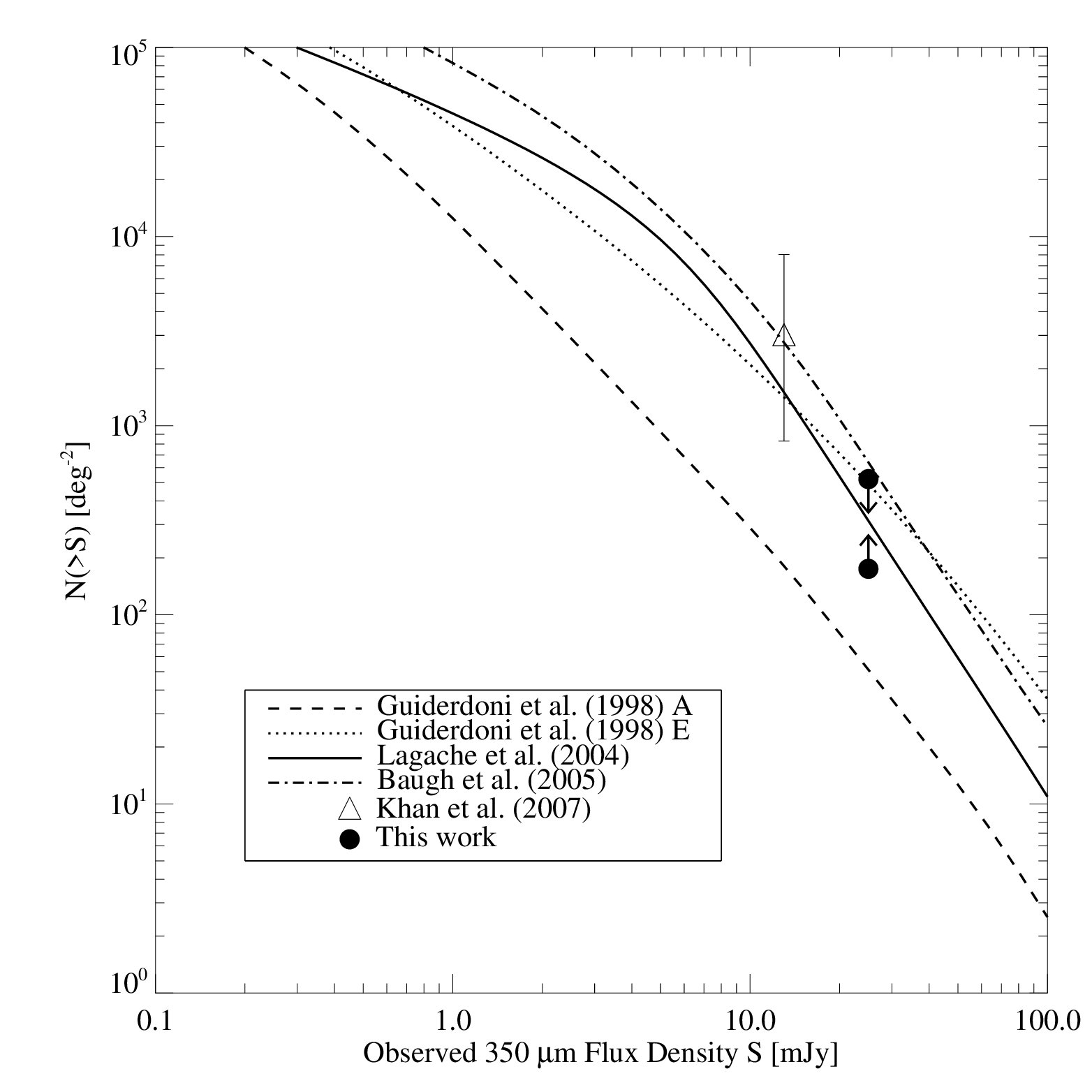,width=0.5\textwidth,scale=0.5}
\caption[Cumulative $350\,\mathrm{\mu m}$ source counts]{Cumulative $350\,\mathrm{\mu m}$ source counts.  The solid line is a prediction of the source counts from \citet{Lagache2004} using a phenomenological model that constrains the IR luminosity function evolution with redshift.  The dot-dash line is the prediction of the $350\,\mathrm{\mu m}$ counts from the semi-analytic models of \citet{Baugh05} which match the $850\,\mathrm{\mu m}$ number counts.  The dotted and dashed lines are semi-analytic predictions of the source counts resulting from evolutionary scenarios E and A, respectively, from \citet{Guiderdoni}, where star-formation proceeds in `burst' mode and includes either no ULIRGs (scenario A) or an increasing fraction of ULIRGs with redshift (scenario E).  The solid circles with arrows are upper and lower limits on the source counts based on this work (see text) and are consistent with the source count predictions of \citet{Lagache2004} and scenario E of \citet{Guiderdoni}.  Scenario A of \citet{Guiderdoni} is inconsistent with the observational limits, as expected since it contains no ULIRGs.  The source count constraint from \citet{Khan} above 13\,mJy is also plotted (open triangle).}
\label{fig:350bf}
\end{figure}

We now attempt to crudely estimate the $350\,\mathrm{\mu m}$ source counts at our typical observed flux density, even though the SHARC-II data are not blank-field observations.  We place an upper limit on the source counts using the number of detections acquired in the observed area; one would expect to do \textit{worse} than this in a blank-field survey, since here known SMGs were observed.  Given a total observed area of $48.3\,\mathrm{arcmin}^{2}$ and seven detections above a $350\,\mathrm{\mu m}$ flux density of $\simeq25$\,mJy, we estimate $N(>\!S)\lesssim500\,\mathrm{deg}^{-2}$.  We estimate a lower limit on the source counts by applying the SHARC-II detection rate of $850\,\mathrm{\mu m}$ sources to the whole SHADES area; one would expect to do at least as well in a blank-field search.  We estimate a lower limit of $N(>\!S)\gtrsim200\,\mathrm{deg}^{-2}$, given a SHARC-II detection success rate of 7/24 and a total of 120 SHADES sources found in $\simeq720\,\mathrm{arcmin}^{2}$.  Note that we have neglected the possibility of enhanced density from clustering.  The limits derived here are consistent with the predictions from \citet{Lagache2004} and scenario E from the semi-analytic models of \citet{Guiderdoni} (see Fig.~\ref{fig:350bf}).  Although crude, this is the best available estimate of the $350\,\mathrm{\mu m}$ source counts at these flux densities.  \citet{Khan} also report the $350\,\mathrm{\mu m}$ source counts above 13\,mJy from a deep survey with SHARC-II.  Large surveys planned with the Balloon-borne Large Aperture Submm Telescope (BLAST; \citealt{Devlin}), \textit{Herschel} \citep{Pilbratt}, SCUBA-2 \citep{Holland2006}, and \textit{Planck} \citep{Planck} will be able to provide further constraints on the source counts at shorter submm wavelengths and over a wider dynamic range of flux densities.

\subsection{SEDs of SMGs}\label{SED}

The SEDs of SMGs is dominated by thermal emission from cold dust.
SHARC-II photometry of the SHADES sources can determine the apparent
temperature of the dust.  In conjunction with knowledge of the
redshift, this allows inference of dust masses, FIR luminosities, and
SFRs, allowing one to place these objects in context
with other populations of high-redshift star-forming galaxies and AGN.

The shape of a luminous dusty galaxy SED is well approximated by a
modified blackbody spectrum described by the dust temperature,
$T_\mathrm{d}$, and dust emissivity $\epsilon \propto \nu^{\beta}$.
where $\beta$ lies in a physically plausible range of 1--2
\citep{Hildebrand1983}.  There is a degeneracy between $T_\mathrm{d}$
and $\beta$ which cannot be disentangled by our data \citep{Blain}, so
we fix $\beta=1.5$, which is consistent with the finding of
\citet{Dunne} for local galaxies and with
the values obtained for carbonite and silicate grains from laboratory
measurements \citep{Agladze}.  This leaves two free parameters to be
fitted: $T_\mathrm{d}$ and the SED normalisation.

In practice we work in terms of the {\it apparent} dust temperature,
$T_\mathrm{A}=T_\mathrm{d}/(1+z)$  and fit the data for each galaxy to
  $A\nu_\mathrm{obs}^{3+\beta}/[\mathrm{exp}(h\,\nu_\mathrm{obs}/k\,T_\mathrm{A})-1]$, 
where several factors of $(1+z)$ have been absorbed into the {\it
apparent} normalisation, $A$.  

Assuming that the redshift is known, the restframe dust temperature
can be recovered and the luminosity of the SMG can be determined by
integrating the SED. In cases where the spectroscopic redshift is
ambiguous, the most reliable or probable spectroscopic redshift available
from the literature has been chosen and  is listed in
Table~\ref{tab:tdm}.  For sources with only a 90 per cent confidence
photometric redshift lower limit, the redshift is taken at this
limit (see Table~\ref{tab:sharc_photom}).

We characterise the shape of the SEDs independently
of the photometric redshifts 
because we do not use the radio data in our fits to the SED, and
where  photometric
redshifts are used they have been determined using all available FIR and radio
photometry (except for the $350\,\mathrm{\mu m}$ data from this paper).
 
More complex SED modelling was not attempted (e.g.,~fitting
two-temperature dust components as in \citealt{Dunne}), since this
would require more free parameters than can be constrained using the
typically 2--3 photometric points which exist for each of our SMGs.
The Wien side of the spectrum is sometimes also modified by a
power-law of the form $S_{\nu} \propto \nu^{-\alpha}$ to account for
the increase in optical depth in this part of the spectrum and to
provide a better fit to observational data (see
\citealt{BlainBarnard}).  We neglect such elaborations here (see also
\citealt{Kovacs}), since the Wien side of the spectrum is not sampled
with our data.  $24\,\mathrm{\mu m}$ photometry are available for
SHADES sources but are complicated to interpret since this band
samples PAH and stellar emission features in high-redshift
star-forming galaxies (e.g.~\citealt{Sajina05}).

\begin{figure*}
\epsfig{file=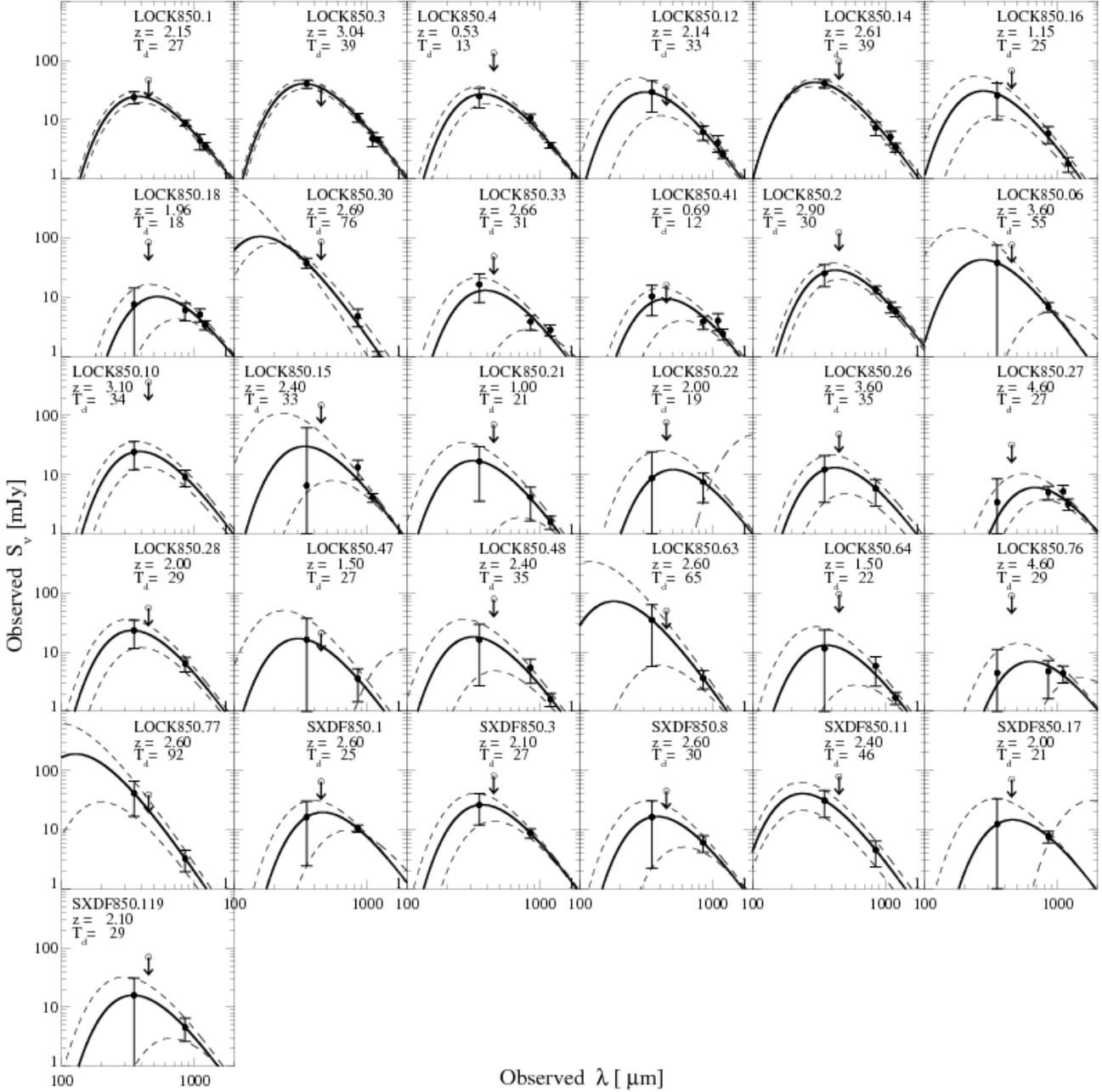,width=1.0\textwidth,scale=0.5}
\caption{
  Best-fitting modified blackbody SEDs for SHADES $850\,\mathrm{\mu
    m}$-selected SMGs with $350\,\mathrm{\mu m}$ photometry with
  $\beta$ fixed to 1.5, in the same order as Table~\ref{tab:tdm}.  The first ten SEDs use the available 
spectroscopic redshifts, and the rest use only photometric redshift constraints.
  The dashed curves show the $1\,\sigma$ range in the best-fit SED.  
  The solid circles
  indicate photometry used to fit the SED, while the open circles show
  the SCUBA $450\,\mathrm{\mu m}$ upper limits which are not used explicitly
  in the fits (though they are never violated).  The spectroscopic or
  photometric redshifts indicated in each
  panel have been used to recover the given restframe $T_\mathrm{d}$.
  Notice that the  $350\,\mathrm{\mu m}$ point
  provides almost all of the constraining power on the dust temperature.}
\label{fig:fits2}
\end{figure*}

The results of this fitting process are plotted in Fig.~\ref{fig:fits2}.  
Notice that the $350\,\mathrm{\mu m}$ data point
provides almost all of the contraining power on $T_\mathrm{A}$.  

\begin{figure}
\epsfig{file=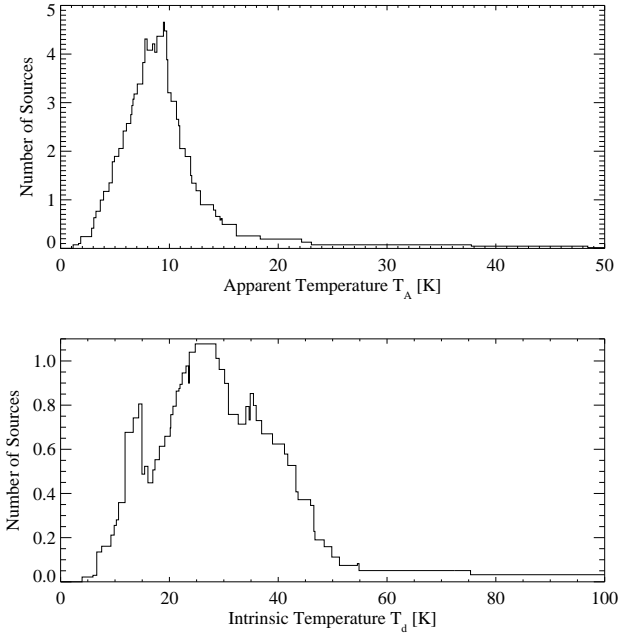,width=0.5\textwidth,scale=0.5}
\caption[]{Apparent (observed) and restframe dust temperature distributions of an unbiased sample of SMGs.  Each distribution has been calculated by adding a rectangle of area one centred on the source temperature ($T_\mathrm{A}$ or $T_\mathrm{d}$), with a width given by the errors in the temperature (see Table~\ref{tab:tdm}) to the histogram. The $T_\mathrm{A}$ distribution reveals that our survey selects SMGs with $T_\mathrm{A}\simeq(8\pm3)$\,K. The $T_\mathrm{d}$ distribution contains bumpy features around 12\,K, 28\,K, and 35\,K and is found to be broader than we would expect by chance (see text).  The underlying distribution is consistent with most SMGs having $T_\mathrm{d}\simeq28$\,K but with a few being a bit hotter ($\simeq35$\,K).}
\label{fig:tdist}
\end{figure}

We have measured the apparent temperatures ($T_\mathrm{A}$) of an unbiased sample of SMGs from SHADES.  We calculate the corresponding temperature distribution by adding a rectangle of area one centred on the source $T_\mathrm{A}$, with a width given by the errors in $T_\mathrm{A}$ to the histogram (see Fig.~\ref{fig:tdist}).  It appears that our survey selects SMGs with $T_\mathrm{A}\simeq(8\pm3)$\,K.  We recover the intrinsic dust temperature ($T_\mathrm{d}$) distribution of SMGs by shifting each SED into the restframe, following a similar procedure to that described for the $T_\mathrm{A}$ distribution above (see Fig.~\ref{fig:tdist}).  The $T_\mathrm{d}$ distribution appears to contain structure (`bumps' centred around 12\,K, 28\,K, and 35\,K) and is broader compared with the $T_\mathrm{A}$ distribution.  To investigate if this distribution is more structured and/or wider than we would expect by chance, we have simulated the temperature distribution we would expect to see if all SMGs have $T_\mathrm{d}=28$\,K.  The simulated $T_\mathrm{d}$ distribution is constructed by adding an error box of area one centred on $T_\mathrm{d}=28\,\mathrm{K}+\mathrm{rand}\times\sigma$, where `rand' is a normally-distributed random number, for each source.  Although the shape and width of the simulated distribution changes notably depending on the random realisation, we \textit{never} reproduce the `bumps' at $T_\mathrm{d}\simeq12$ or 35\,K and we \textit{always} predict more SMGs at a few K than we see.  Therefore we conclude that the underlying distribution is just a bit wider than we expect by chance and is consistent with most SMGs at 28\,K with a few hotter ones.  Now we compare the dust temperatures of our photometric and spectroscopic redshift sub-samples and to previous estimates for SMGs.

We derive a mean $T_\mathrm{d}=(31\,\pm{18})\,\mathrm{K}$ (median $T_\mathrm{d}=31\,\mathrm{K}$) for our sample of SMGs with spectroscopic redshifts.  For the sample of 21 $350\,\mathrm{\mu m}$-observed SMGs with only photometric redshifts, the mean is $T_\mathrm{d}=(35\,\pm{17})\,\mathrm{K}$ (median $T_\mathrm{d}=29\,\mathrm{K}$).  A two-sided KS test reveals that the temperature distributions of the photometric redshift and spectroscopic redshift subsets are consistent with being drawn from the same distribution.  Therefore we quote a mean $T_\mathrm{d}=(34\,\pm{17})\,\mathrm{K}$ (median $T_\mathrm{d}=29\,\mathrm{K}$) for our entire sample of SMGs.  Our mean $T_\mathrm{d}$ is consistent with previous estimates for SMGs from \citet{Chapman2005} ($T_\mathrm{d}=36\,\mathrm{K}$ for radio-identified SMGs spectroscopic redshifts), from \citet{Kovacs} ($T_\mathrm{d}=35\,\mathrm{K}$), and from \citet{Pope2006} ($T_\mathrm{d}\simeq30\,\mathrm{K}$ for SMGs in the Great Observatories Origins Deep Survey-North, GOODS-N, region).  In contrast, we find that the \citealt{Farrah} ULIRG sample has a median $T_\mathrm{d}=42\,\mathrm{K}$ when their photometry are fit with our greybody (see also \citealt{Clements08}).  Thus SMGs appear to have lower temperatures on average than local ULIRGs, although bear in mind that local ULIRGs are selected at $60\,\mathrm{\mu m}$ which inherently introduces a bias to selecting hotter objects than submm observations would typically find.

The SED is integrated to estimate the bolometric luminosity of the
SMG.  Assuming that the FIR luminosity is predominantly powered by
star-formation (i.e.~negligible contribution from an AGN), coming from
a starburst of less than 100\,Myr, with a \citet{Salpeter55} initial
mass function, then the SFR can be calculated following
\citet{Kennicutt}:

\begin{equation}\label{SFR}
\mathrm{SFR}(\mathrm{M}_{\odot}\,\mathrm{yr}^{-1})=1.7\times10^{-10}\,L_\mathrm{FIR}(\mathrm{L_\odot}).
\end{equation}

The median FIR luminosity for our sample is $\simeq2\times10^{12}\,\mathrm{L}_{\odot}$ (the mean is $\simeq8\times10^{12}\,\mathrm{L}_{\odot}$, with a large scatter), which is consistent with what has been found previously for SMGs (e.g.~\citealt{Kovacs}).  The median SFR for the $350\,\mathrm{\mu m}$-observed SHADES sources is $\simeq400\,\mathrm{M}_{\odot}\,\mathrm{yr}^{-1}$, as calculated from Equation~\ref{SFR}.  This is consistent with previous work, indicating that SMGs are significant contributors in the global star-formation of the high-redshift Universe (see e.g.~\citealt{Lilly99}; \citealt{Blain}; \citealt{Chapman2005}; \citealt{paper4}).  

\begin{table*}
\begin{minipage}{1.0\textwidth}
\scriptsize
\caption
[] 
{Derived dust temperatures ($T_\mathrm{d}$), FIR luminosities
  ($L_\mathrm{FIR}$), illuminated dust masses ($M_\mathrm{d}$) and
  SFRs for the $350\,\mathrm{\mu m}$-observed SHADES SMGs.  The
  emissivity, $\beta$, is fixed at 1.5 and the best spectroscopic or
  photometric redshifts are used in fitting to the 350 and
  $850\,\mathrm{\mu m}$ photometry, plus 1.1 and 1.2\,mm data if
  available.  Errors in the fit parameters (i.e.~the position of the
  peak and the SED normalisation) are derived from the 68 per cent
  $\chi^{2}$ confidence interval around the best-fitting values and
  have been carried through to all the derived quantities.  Where
  appropriate, errors in the photometric redshifts have also been folded
  into the uncertainties.}
\label{tab:tdm}
\begin{tabular}{lcccccc}
\hline 
\multicolumn{1}{c}{SHADES ID} & \multicolumn{1}{c}{$z$} & \multicolumn{1}{c}{$T_\mathrm{d}$} & \multicolumn{1}{c}{$L_\mathrm{FIR}$} & \multicolumn{1}{c}{$M_\mathrm{d}$} & \multicolumn{1}{c}{SFR} & \multicolumn{1}{c}{Notes}\\
\multicolumn{1}{c}{} & \multicolumn{1}{c}{} & \multicolumn{1}{c}{(K)} & \multicolumn{1}{c}{($\times10^{12}\,\mathrm{L}_{\odot}$)} & \multicolumn{1}{c}{($\times10^{9}\,\mathrm{M}_{\odot}$)} & \multicolumn{1}{c}{($\mathrm{M}_{\odot}\,\mathrm{yr}^{-1}$)}& \multicolumn{1}{c}{}\\
\hline 
With a spectroscopic redshift & & & & & & \\
LOCK850.01 & 2.148 & $27^{+4}_{-3}$ & $2.3^{+1.0}_{-0.8}$ & $1.4^{+0.6}_{-0.3}$ & $390^{+170}_{-130}$ & see \citet{Kovacs} \\
LOCK850.03 & 3.036 & $39^{+4}_{-4}$ & $9.6^{+3.5}_{-2.8}$ & $0.8^{+0.2}_{-0.2}$ & $1600^{+600}_{-480}$ & see \citet{Kovacs}\\
LOCK850.04 & 0.526 & $13^{+2}_{-2}$ & $0.08^{+0.03}_{-0.03}$ & $2.7^{+0.9}_{-0.6}$ & $13^{+6}_{-4}$ & \\
LOCK850.12 & 2.142 & $33^{+14}_{-12}$ & $3.2^{+8.2}_{-2.6}$ & $0.7^{+1.4}_{-0.3}$ & $550^{+1400}_{-440}$ & see \citet{Kovacs}\\
LOCK850.14 & 2.611 & $39^{+7}_{-5}$ & $8.0^{+4.6}_{-2.8}$ & $0.5^{+0.2}_{-0.1}$ & $1400^{+790}_{-480}$ & see \citet{Kovacs} \\ 
LOCK850.16 & 1.147 & $25^{+9}_{-7}$ & $0.8^{+1.6}_{-0.5}$ & $0.9^{+0.8}_{-0.4}$ & $140^{+280}_{-90}$ & \\
LOCK850.18 & 1.956 & $18^{+5}_{-7}$ & $0.5^{+0.6}_{-0.4}$ & $2.7^{+17.0}_{-1.3}$ & $90^{+100}_{-70}$ & see \citet{Kovacs} \\
LOCK850.30 & 2.689 & $76^{+103}_{-21}$ & $40.0^{+932.0}_{-26.0}$ & $0.1^{+0.1}_{-0.1}$ & $6900^{+160000}_{-4500}$ & see \citet{Kovacs} \\
LOCK850.33 & 2.664 & $31^{+8}_{-13}$ & $1.9^{+2.1}_{-1.5}$ & $0.5^{+2.1}_{-0.2}$ & $330^{+370}_{-270}$ & \\
LOCK850.41 & 0.689 & $12^{+4}_{-6}$ & $0.04^{+0.05}_{-0.03}$ & $1.7^{+12.0}_{-0.8}$ & $7^{+9}_{-6}$ & see \citet{Kovacs} \\
\\ 
\hline
With a photometric redshift & & & & & & \\
LOCK850.02 & $2.9^{+0.7}_{-0.1}$ & $30^{+12}_{-8}$ & $4.8^{+8.3}_{-2.9}$ & $1.9^{+2.0}_{-1.0}$ & $830^{+1430}_{-490}$ & \\
LOCK850.06 & $3.6^{+1.0}_{-0.1}$ & $55^{+51}_{-33}$ & $19.0^{+930.0}_{-19.0}$ & $0.3^{+2.5}_{-0.2}$ & $3300^{+160000}_{-3200}$ & $\chi^2\ll1$ \\
LOCK850.10 & $3.1^{+0.9}_{-0.3}$ & $35^{+17}_{-10}$ & $5.3^{+4.5}_{-3.5}$ & $0.9^{+1.2}_{-0.3}$ & $920^{+780}_{-610}$ & $\chi^2\ll1$\\
LOCK850.15 & $2.4^{+0.4}_{-0.4}$ & $33^{+17}_{-29}$ & $4.0^{+37.0}_{-3.6}$ & $1.5^{+9.0}_{-1.0}$ & $690^{+6400}_{-630}$ & \\
LOCK850.21 & $\geq1.0$ & $>21^{+8}_{-7}$ & $>0.3^{+0.6}_{-0.2}$ & $>0.8^{+1.5}_{-0.4}$ & $>60^{+100}_{-40}$ & \\
LOCK850.22 & $\geq2.0$ & $>19^{+10}_{-10}$ & $>0.7^{+18.0}_{-0.7}$ & $>2.9^{+49.0}_{-1.8}$ & $>120^{+3000}_{-110}$ & $\chi^2\ll1$\\
LOCK850.26 & $3.6^{+0.1}_{-0.8}$ & $35^{+11}_{-15}$ & $3.8^{+30.0}_{-3.1}$ & $0.5^{+1.7}_{-0.3}$ & $650^{+5200}_{-540}$ & $\chi^2\ll1$\\
LOCK850.27 & $4.6^{+1.4}_{-0.4}$ & $27^{+22}_{-11}$ & $1.9^{+1.2}_{-1.2}$ & $1.2^{+9.1}_{-1.0}$ & $330^{+1200}_{-210}$ & \\
LOCK850.28 & $\geq2.0$ & $>29^{+7}_{-7}$ & $>2.0^{+1.3}_{-1.3}$ & $>0.9^{+0.9}_{-0.9}$ & $>340^{+220}_{-220}$ & $\chi^2\ll1$\\
LOCK850.47 & $\geq1.5$  & $>27^{+19}_{-19}$ & $>0.8^{+41.0}_{-0.8}$ & $>0.5^{+37.0}_{-0.3}$ & $>140^{+7100}_{-140}$ & $\chi^2\ll1$\\
LOCK850.48 & $2.4^{+0.5}_{-0.1}$ & $35^{+20}_{-12}$ & $2.6^{+9.4}_{-2.0}$ & $0.5^{+0.8}_{-0.3}$ & $440^{+1600}_{-340}$ & \\
LOCK850.63 & $2.6^{+0.4}_{-0.4}$ & $65^{+71}_{-59}$ & $22.0^{+2600.}_{-22.0}$ & $0.1^{+850.0}_{-0.08}$ & $3800^{+460000}_{-3800}$ & $\chi^2\ll1$ \\
LOCK850.64 & $\geq1.5$ & $>22^{+8}_{-8}$ & $>0.5^{+0.9}_{-0.4}$ & $>1.3^{+2.8}_{-0.6}$ & $>90^{+150}_{-70}$ & \\
LOCK850.76 & $4.6^{+1.4}_{-1.1}$ & $29^{+21}_{-19}$ & $2.4^{+2.1}_{-2.0}$ & $0.9^{+100.0}_{-0.4}$ & $410^{+360}_{-350}$ & \\
LOCK850.77 & $2.6^{+0.8}_{-0.1}$ & $92^{+140}_{-75}$ & $82.0^{+13000.0}_{-82.0}$ & $0.06^{+0.6}_{-0.4}$ & $14000^{+2300000}_{-14000}$ & $\chi^2\ll1$, unphysical, misID? \\
\\
SXDF850.1 & $2.6^{+0.4}_{-0.3}$ & $25^{+10}_{-8}$ & $2.3^{+0.8}_{-1.8}$ & $2.2^{+4.4}_{-1.2}$ & $400^{+130}_{-300}$ & $\chi^2\ll1$\\
SXDF850.3 & $2.1^{+0.3}_{-0.1}$ & $27^{+10}_{-8}$ & $2.3^{+1.7}_{-1.6}$ & $1.4^{+1.5}_{-0.6}$ & $400^{+300}_{-270}$ & $\chi^2\ll1$\\
SXDF850.8 & $2.6^{+1.3}_{-0.1}$ & $30^{+11}_{-11}$ & $2.4^{+3.9}_{-2.0}$ & $0.8^{+2.0}_{-0.4}$ & $410^{+680}_{-340}$ & $\chi^2\ll1$\\
SXDF850.11 & $2.4^{+0.4}_{-0.4}$ & $46^{+29}_{-24}$ & $7.7^{+190.0}_{-7.3}$ & $0.3^{+0.9}_{-0.1}$ & $1300^{+32000}_{-1300}$ & $\chi^2\ll1$\\
SXDF850.17 & $\geq2.0$ & $>21^{+11}_{-11}$ & $>0.9^{+23.0}_{-0.9}$ & $>2.2^{+36.0}_{-1.4}$ & $>160^{+3900}_{-150}$ & $\chi^2\ll1$\\
SXDF850.119 & $2.1^{+0.1}_{-0.1}$ & $29^{+15}_{-14}$ & $1.5^{+27.0}_{-1.4}$ & $0.6^{+2.8}_{-0.3}$ & $260^{+4700}_{-240}$ & $\chi^2\ll1$\\
\hline
\end{tabular}
\end{minipage}
\end{table*}
\normalsize

We can also constrain the cold dust mass,
$M_\mathrm{d}$, in these obscured star-forming galaxies. 
The flux density of a galaxy at an observed frequency, $\nu_\mathrm{obs}$, is given
by the usual relation (e.g.~\citealt{HDR97}):

\begin{equation}\label{mass}
S_{\nu_\mathrm{obs}}=B_{\nu'}(T)\kappa_{\nu'}\,M_\mathrm{d}\,(1+z)/D_{\mathrm{L}}^{2},
\end{equation}

\noindent 
where $D_{\mathrm{L}}$ is the cosmological luminosity distance and
$B_{\nu'}$ is the Planck function evaluated at the emitted frequency,
$\nu'=\nu_\mathrm{obs}\,(1+z)$.  The quantity $\kappa_{\lambda}$ is the
wavelength-dependent mass-absorption coefficient (or `effective area'
for blackbody emission by a certain mass of dust) and
$\kappa_{\lambda}$ can be  extrapolated from an average 
$\kappa_{125_\mathrm{\mu
    m}}=2.64\,\mathrm{m}^{2}\,\mathrm{kg}^{-1}$ 
assuming $\beta=1.5$ \citep{Dunnekappa}.

Using Equation~\ref{mass}, the median dust mass implied by the $850\,\mathrm{\mu m}$ observations is $9\,\times10^{8}\,\mathrm{M}_{\odot}$.  Uncertainties in the dust masses are dominated by the uncertainty in the photometric redshifts (when these are used), $T_\mathrm{d}$ and the flux density at $850\,\mathrm{\mu m}$.  The uncertainty in $\kappa_{\nu}$ is a factor of a few and has not been included in the dust mass error bars, but the relative dust masses in our sample will be correct if the same value holds for all SMGs (see \citealt{Blain} and references therein for a discussion).  Assuming that the maximum possible interstellar dust mass for a galaxy is about 1/500 of its total baryonic mass (see \citealt{EdmundsEales}), the total baryonic mass of each SMG is estimated to be $M_\mathrm{bary}\simeq\!5\times10^{11}\,\mathrm{M}_{\odot}$.  For comparison, the baryonic mass of the Milky Way is about $10^{11}\,\mathrm{M}_{\odot}$.

The best-fitting SEDs are given in Fig.~\ref{fig:fits2}, while
Table~\ref{tab:tdm} gives the best-fitting value of $T_\mathrm{d}$ for
each SMG, as well as the derived $L_\mathrm{FIR}$, $M_\mathrm{d}$ and
SFR. 

Errors in the fit parameters (i.e.~the position of the peak and the
SED normalisation) are derived from the 68 per cent $\chi^{2}$
confidence interval around the best-fitting values.  These errors have
been carried through to all the other derived quantities in
Table~\ref{tab:tdm}.  Obviously, for galaxies with only two FIR
photometric points, the SEDs are not well-constrained and the minimum
reduced $\chi^{2}$ values are significantly less than one.
Photometric redshift errors have been folded into the uncertainties
for derived quantities.  We assume that our adopted model (a 
greybody with $\beta=1.5$) is correct and so the quoted errors do not include a contribution from possible systematic errors.  Adopting a
constant value of $\beta$ is equivalent to assuming that the dust has
similar properties in all galaxies.  This is probably a reasonable
assumption.  As additional precise photometric data become available (from
\textit{Herschel} for example) $\beta$ will be able to be constrained
by direct fitting.  Note, however, that changing the value of $\beta$
by $\pm{0.5}$ (between physically plausible values) has the effect of
changing the derived dust temperature by about $\pm5$--10\,K, which in
turn affects the derived luminosities of the SMGs by $\pm20$--40 per
cent.

The SMG population is therefore confirmed to be dominated by massive ($>10^{12}\,\mathrm{M}_{\odot}$, assuming $M_\mathrm{tot}/M_\mathrm{baryons}\simeq6$), luminous ($\simeq\!2\times10^{12}\,\mathrm{L}_{\odot}$) star-forming (SFR $\simeq400\,\mathrm{M}_{\odot}\,\mathrm{yr}^{-1}$) galaxies with dust temperatures of $\simeq35\,\mathrm{K}$.  In contrast, local starburst galaxies with similar temperatures (\citealt{Dunne2000}: $\bar{T}_\mathrm{d}=36$\,K) are not usually classified as ULIRGs, since they are about ten times less luminous ($L_\mathrm{FIR}\sim10^{11}\,\mathrm{L}_{\odot}$).  In addition, SMGs have more than ten times more dust than local starburst galaxies, which could be explained by any of the following reasons:  (1) SMGs are hosted in galaxies ten times more massive than local starbursts; (2) SMGs are more gas- and dust- rich for a given baryonic mass, being at an earlier stage in evolution with a higher gas fraction; or (3) the dust properties evolve with redshift and high-redshift dust has a higher emissivity and so produces more submm emission for a given mass of dust.

\section{Discussion}\label{discussion}

\subsection{Trends in the data and consideration of selection effects}

\begin{figure}
\epsfig{file=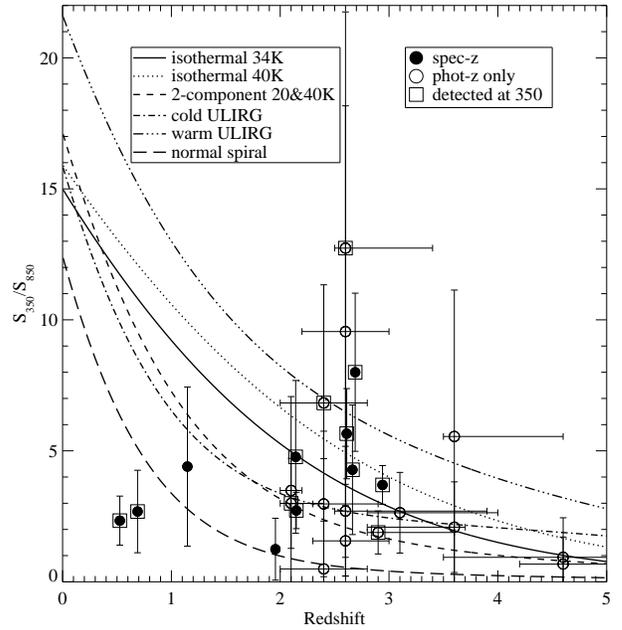,width=0.5\textwidth,scale=0.5}
\caption{Observed $S_{350}/S_{850}$ colour versus redshift 
using the values from Table~\ref{tab:sharc_photom}.  See 
inset for a description of the plot symbols.  Sources 
with only redshift lower limits are omitted from this 
plot.  The colour appears to be independent of redshift 
in these data (the median colour is 3.0), which is not 
surprising given the lack of dynamic range in the flux 
densities of the detected sources at $850\,\mathrm{\mu m}$.  
Given a higher dynamic range in flux in future studies, 
we could attempt to trace SED evolution with redshift by 
distinguishing between models.  We have plotted some 
plausible models including (see inset for legend) an 
isothermal greybody with $T_\mathrm{d}=40$\,K, 
$\beta=1.5$ (e.g.~\citealt{Blain}); an isothermal 
greybody $T_\mathrm{d}=34$\,K, $\beta=1.5$ (this work); 
a 2-component typical IR-bright starburst galaxy with 
$T_\mathrm{d}=20$ \& 40\,K, $\beta=2$, and 
$M_\mathrm{cold}/M_\mathrm{warm}=30$ (e.g.~\citealt{Dunne}); 
a theoretical warm ULIRG $T_\mathrm{d}=24$ \& 50\,K, $\beta=2$, 
and $M_\mathrm{cold}/M_\mathrm{warm}=5$; a normal spiral 
(e.g.~Milky Way, NGC891, or M51) with 
$T_\mathrm{d}=15$ \& 30\,K, $\beta=2$, and 
$M_\mathrm{cold}/M_\mathrm{warm}=100$; a cold 
ULIRG (e.g.~Arp220) with $T_\mathrm{d}=18$ \& 48\,K, 
$\beta=2$, and $M_\mathrm{cold}$/$M_\mathrm{warm}=42$ .
}\label{fig:colour}
\end{figure}

\begin{figure}
\epsfig{file=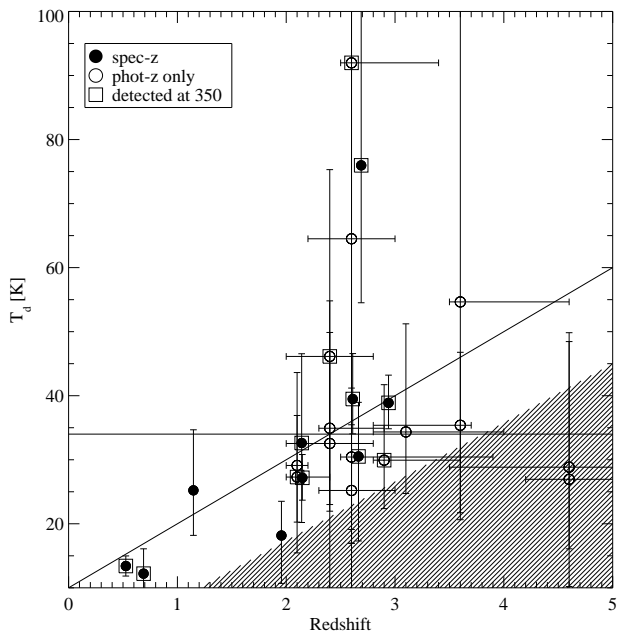,width=0.5\textwidth,scale=0.5}
\caption{Derived dust temperature versus redshift for the sample of SMGs.  See inset for a description of the plot symbols.  The horizontal line represents the median dust temperature of the sample.  The diagonal solid line indicates an observed dust temperature of 10\,K, i.e.~$T_\mathrm{d}=10\,(1+z)\,\mathrm{K}$, and appears to fit the data better; a similar conclusion was also noted by \citet{Kovacs}.  This is not surprising, given that we find a constant $S_{350}/S_{850}$ with redshift, as seen in Fig.~\ref{fig:colour}. Combined with Monte Carlo simulations of the sample selection effects (see text), it appears that the absence of 30\,K sources seen at $z\gtrsim3$ is mainly due to a selection effect from SHARC-II, indicated by the shaded region.  The lack of objects in the upper left is the result of evolution and shows that the $850\,\mathrm{\mu m}$ survey detects fewer warm SMGs in the low-redshift Universe than at high-redshift.  See also \citet{Chapman2005}.}\label{fig:tvsz}
\end{figure}

We first consider selection effects in our data before looking for trends in the derived properties.  It is not surprising that the $S_{350}/S_{850}$ colour appears to be independent of redshift (see Fig.~\ref{fig:colour}), owing to the small dynamic range in flux densities of the $850\,\mathrm{\mu m}$-selected sources.  The colour could evolve with redshift, given a wider dynamic range.  With this in mind, the relationship between dust temperature and redshift is examined in Fig.~\ref{fig:tvsz}.  Notice that the spectroscopic and photometric redshift samples have about the same scatter and that the data deviate about the horizontal line in Fig.~\ref{fig:tvsz} (the sample median) by about $\pm{5}\,\mathrm{K}$ between redshifts of about 1 and 4 (except for a few outliers).  It appears that the SMGs span a range of dust temperatures, with an apparent trend of $T_\mathrm{d}$ increasing with $z$.  A Spearman rank test reveals a strong correlation in the data with spectroscopic redshifts and a weaker correlation when the photometric redshift sample is included.  Is this due to selection effects or is this evidence for evolution of SMGs/ULIRGs with redshift?  

In order to test for selection effects in our data, we populate the $T_\mathrm{d}$-$z$ parameter space of Fig.~\ref{fig:tvsz} by assuming reasonable distributions of SMG properties and apply 850 and $350\,\mathrm{\mu m}$ cuts to the data to see if the trends seen in Fig.~\ref{fig:tvsz} can be reproduced by selection effects alone.  The dust temperature distribution of SMGs is assumed to be a simple Gaussian with a mean of $35\,\mathrm{K}$ and $\sigma=6\,\mathrm{K}$, consistent with observations (e.g.~\citealt{Kovacs}; \citealt{Huynh07}).  We create a $\beta=1.5$ greybody SED with a temperature randomly selected from the temperature distribution, with a normalisation constrained so that $L_\mathrm{FIR}$ is within the range of what we know about SMGs, i.e.~a simple Gaussian distribution with a mean of $6.7\,\times10^{12}\,\mathrm{L}_\odot$ and $\sigma=3.0\,\times10^{12}\,\mathrm{L}_\odot$ (e.g.~\citealt{Pope}).  A redshift for each source is randomly selected from a simple uniform redshift distribution spanning $0\leq z \leq 5$ to allow observed 850 and $350\,\mathrm{\mu m}$ fluxes to be calculated.  We reject an SMG from our surveyed sample if its 850 and $350\,\mathrm{\mu m}$ flux densities are less than 4\,mJy and 25\,mJy, respectively.  This procedure is repeated 100 times and the shaded region in Fig.~\ref{fig:tvsz} indicates the region where selection effects preclude the detection of SMGs in our $350\,\mathrm{\mu m}$ survey of $850\,\mathrm{\mu m}$-selected SMGs.  We miss sources in the lower right-hand region (high-redshift, low temperature) of Fig.~\ref{fig:tvsz} because they are not bright enough at $850\,\mathrm{\mu m}$ to make it into our SHADES sample.  In addition some of the $850\,\mathrm{\mu m}$-selected sources are too faint at $350\,\mathrm{\mu m}$.  On the other hand, there seems to be a dearth of sources in the data at low redshifts with high dust temperatures (and consequently, high luminosities), where we are not prone to any selection effects (Fig.~\ref{fig:tvsz}).  The lack of sources in the upper left of Fig.~\ref{fig:tvsz} is probably a consequence of the long-established strong luminosity evolution of the submm population (see fig.~2 in \citealt{Smail97}), i.e.~there are not as many sources at low redshift with temperatures $\gtrsim 35\,\mathrm{K}$ with sufficient luminosity to be detected by SHADES at $850\,\mathrm{\mu m}$.  Of course there could still be a systematic temperature dependence which might be seen by deep 450 and $850\,\mathrm{\mu m}$ SCUBA-2 data through the detection of many more $450\,\mathrm{\mu m}$ sources at fainter flux densities. 
\begin{figure}
\epsfig{file=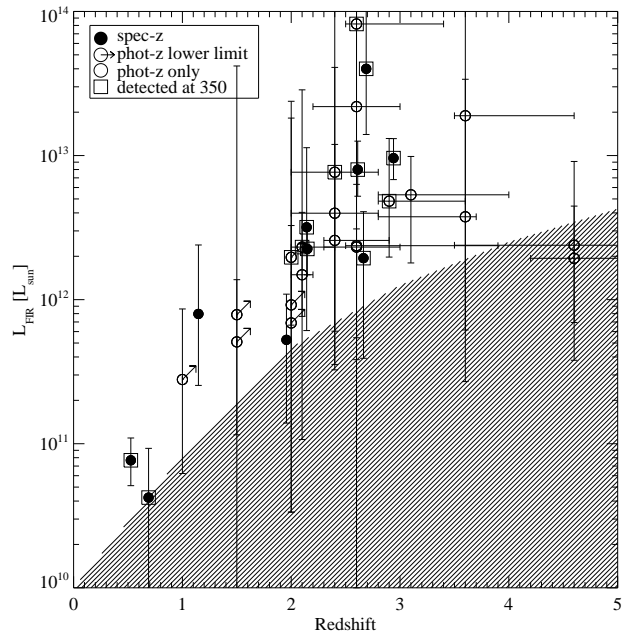,width=0.5\textwidth,scale=0.5}
\caption[Far-IR luminosity versus redshift]{Bolometric luminosity versus redshift for the sample of SMGs. The plot symbols are the same as in Fig.~\ref{fig:colour}, except that arrows now indicate points with redshift lower limits.  The shaded region indicates selection effects from SHARC-II.  There appears to be a dearth of higher luminosity sources at low redshift:  a hint of a strong luminosity evolution for SMGs with redshift.}\label{fig:lfirvsz}
\end{figure}

This idea can be further investigated when the relationship between $L_\mathrm{FIR}$ and redshift is examined (Fig.~\ref{fig:lfirvsz}), since there is an intrinsic correlation between temperature and $L_\mathrm{FIR}$.  As before, SHARC-II is insensitive to the shaded region in Fig.~\ref{fig:lfirvsz} (i.e.~misses SMGs cooler than $T_\mathrm{d}/(1+z)\simeq$7\,K).  Due to the increased volume at low redshifts, we would expect to see about two high-luminosity sources between redshifts of 0 and 1.  Thus there appears to be a hint of a dearth of higher luminosity SMGs at the lowest redshifts compared to higher redshifts that cannot be explained by selection or volume effects.  

These results are difficult to make sense of unless a strong SMG evolution with redshift is invoked.  Millimetre-wave cameras such as MAMBO, Bolocam and AzTEC might be sensitive to cooler SEDs and could perhaps select SMGs to fill in this region of the ($L_\mathrm{FIR}$, $z$) plane.  There also seems to be a lack of SMGs at low redshifts and high luminosity, indicating that SCUBA-selected SMGs are intrinsically more luminous at high redshifts.  This is consistent with what was found by \citet{Ivison} and \citet{Pope2006} and suggests a strong evolution of SMGs/ULIRGs with redshift.  Thus, Fig.~\ref{fig:tvsz} is just a reflection of the known effect seen in Fig.~\ref{fig:lfirvsz}, i.e.~that the most luminous SMGs are at higher redshifts.  This means that they must be evolving very steeply with redshift (i.e.~`down-sizing').  This idea has already been explored in \citet{Wall}.  Larger samples of SMGs with robust redshifts and a wider dynamic range in 350 and $850\,\mathrm{\mu m}$ flux densities and luminosity are needed to test this idea further and to fully describe the evolution of the luminosity function of SMGs.

\section{Conclusions}\label{fp}

We have performed follow-up mapping at $350\,\mathrm{\mu m}$ using SHARC-II of several $850\,\mathrm{\mu m}$-selected sources from SHADES.  In total there are flux density measurements for about 25 per cent of the SHADES SMGs, of which this work has provided 21 new flux density constraints. 

The combination of 350 and $850\,\mathrm{\mu m}$, 1.1, and/or 1.2\,mm photometry and spectroscopic or photometric redshifts have provided estimates of $T_\mathrm{d}$ and $L_\mathrm{FIR}$ for each $350\,\mathrm{\mu m}$-observed SHADES SMG by fitting a modified blackbody function to the data.  The SMGs population is confirmed to be dominated by very dusty ($M_\mathrm{dust}\simeq9\times10^{8}\,\mathrm{M}_{\odot}$), luminous ($L_\mathrm{FIR}\simeq2\times10^{12}\,\mathrm{L}_{\odot}$) star-forming ($\mathrm{SFR}\simeq400\,\mathrm{M}_{\odot}\,\mathrm{yr}^{-1}$) galaxies with dust temperatures of $\simeq35\,\mathrm{K}$.  We have measured the temperature distribution of SMGs and find that the underlying distribution is slightly broader than implied by the error bars, and that most SMGs are at 28\,K with a few hotter.

With the data in hand, it is very difficult to understand whether we are just seeing a combination of selection effects and the known luminosity evolution of SMGs, or whether the SEDs might also systematically change with redshift.  Larger samples of SMGs with a broader selection function will enable this to be fully investigated, e.g.~with a deep $450\,\mathrm{\mu m}$ survey, a primary goal of the upcoming SCUBA-2 Cosmological Legacy Survey\footnote{{\tt http://www.jach.hawaii.edu/JCMT/surveys/Cosmology.html}}.

Using these data, we have also been able to estimate the $350\,\mathrm{\mu m}$ source counts at $200\,\mathrm{deg}^{-2}\lesssim N(\gtrsim25\,\mathrm{mJy})\lesssim500\,\mathrm{deg}^{-2}$.  Large blank-field surveys at 350 and/or $450\,\mathrm{\mu m}$ will allow us to study a statistically significant population of lower redshift and/or hotter dust temperature galaxies, responsible for emission nearer to the peak of the extragalactic background than those selected at $850\,\mathrm{\mu m}$.  These studies should provide a link between galaxies detected at $\sim\!1\,\mathrm{mm}$ and those selected in the mid-IR with \textit{Spitzer}.  Such comparisons should allow us to investigate the evolution of FIR emitting galaxies and how they contribute to the background at different wavelengths.  

\section{Acknowledgments}\label{ack}

We thank Darren Dowell and Attila Kovacs at Caltech, who provided support and advice before, during, and after the CSO observing runs and to the CSO crew for support.  The CSO is operated by Caltech under a contract from the National Science Foundation (NSF).  KC, MH, AP and DS acknowledge support from the Natural Sciences and Engineering Research Council of Canada (NSERC).  KC also acknowledges the Particle Physics Association Research Council (PPARC) for support.  IRS acknowledges support from the Royal Society.  IA and DH acknowledge support from CONACyT grants.  JW is grateful for support provided by the Max-Planck Society and the Alexander von Humboldt Foundation.  EvK acknowledges support from FWF grant P18493.

\setlength{\bibhang}{2.0em}

\begin{appendix}

\section{Effective exposure times}\label{effective}
When describing astronomical data, it is common and straightforward to give total observation times, or perhaps in chopped systems the total time \textit{on source}.  These times give an indication of the amount of effort expended to collect the data.  However, because the atmosphere is far from transparent at submm wavelengths (even in the atmospheric windows) and the transmission is so weather dependent, the total exposure time does not give a good indication of data quality or expected noise level in the final output.  It is more useful to refer to the \textit{effective total exposure} which is the noise-squared weighted sum of exposure times referred to a completely transparent atmosphere.

The optical depth of the atmosphere in a given direction is determined from the wavelength dependent optical depth at zenith, $\tau_{\lambda}$, and the cosecant of the zenith angle, $\theta$, of observation, called the airmass, $A$.  Thus, the strength of an observed astronomical flux density, $S_\mathrm{o}$ is:

\begin{equation}\label{opticaldepth}
S_\mathrm{o} = S\,e^{-\tau_{\lambda}(t) A(\theta)} .
\end{equation}

At $850\,\mathrm{\mu m}$, the zenith optical depth, $\tau_\mathrm{850}$, ranges from, say, 0.16 in very good weather to 0.7 in bad weather, and the optical depth at 350 or $450\,\mathrm{\mu m}$ is typically 7 times higher.  The first step in combining data collected at different times of night with different values of $\tau_{\lambda}$ and $A$ is to correct $S_\mathrm{o}$ for atmospheric absorption by multiplying by $e^{\tau_{\lambda}\,A}$ (see \citealt{arch01}).

After correcting for atmospheric transients, etc., but not opacity, the \textit{output} noise at the detector is fairly independent of atmospheric opacity, at least for SCUBA operating on the JCMT on Mauna Kea at 450 and $850\,\mathrm{\mu m}$, and for SHARC-II operating at $350\,\mathrm{\mu m}$ at the CSO also on Mauna Kea.  Therefore, the \textit{effective} noise of an observation is inflated exponentially in $\tau_{\lambda}$ and $A$.  The sum

\begin{equation}\label{effectiveeqn}
t_\mathrm{eff} = \sum \Delta t_\mathrm{o} e^{2\,\tau_{\lambda}(t) A}
\end{equation}

\noindent is the integration time that would be required with a completely transparent atmosphere to get the same noise level as a given noise-squared weighted sum of observations, each of size $\Delta\,t_\mathrm{o}$.  We call this the \textit{effective exposure time} and find that tracking this quantity during observations is an effective way to monitor experimental progress.

In practice, the optical depth as a function of time is obtained from automated measurements made at the CSO at $225\,\mathrm{GHz}$ every 10 minutes.  The optical depths at submm wavelengths are very well correlated with these values and one multiplies $\tau_\mathrm{225\,GHz}$ by about 4 and by 24 to get optical depths for SCUBA at 850 and $450\,\mathrm{\mu m}$, respectively, and by 26 to get $\tau$ for SHARC-II at $350\,\mathrm{\mu m}$ \citep{arch01}.

As a numerical example, observation of a source which rises through transit with an average airmass of 1.3 and $\tau_\mathrm{225\,GHz}=0.05$ for 4 hours results in effective exposure times of 2.4 hours at $850\,\mathrm{\mu m}$, 20.7 \textit{minutes} at $450\,\mathrm{\mu m}$ with SCUBA, and 9.5 minutes with SHARC-II.  The shorter effective times at shorter wavelengths are a consequence of the fact that the atmosphere is not very transparent at those wavelengths. This is why space experiments like \textit{Herschel}-SPIRE and \textit{BLAST} are useful even with comparatively small apertures, and why neither of those experiments carries an $850\,\mathrm{\mu m}$ channel.  Experience has shown that effective exposures of between 500 and 600 seconds with SHARC-II are often sufficient to detect faint extragalactic high-redshift sources.

\end{appendix}

\end{document}